\documentclass[preprints,article,accept,moreauthors,pdftex]{Definitions/mdpi}

\firstpage{1} 
\pubvolume{xx}
\issuenum{1}
\articlenumber{5}
\pubyear{2020}
\copyrightyear{2020}
\history{Received: date; Accepted: date; Published: date}

\usepackage{bm} %
\usepackage{verbatim}
\usepackage{xr}
\usepackage{tabularx}
\usepackage{multicol}
\usepackage{lipsum}
\usepackage{ifthen}
\usepackage{subcaption}
\usepackage[normalem]{ulem}
\newboolean{index-notation}
\setboolean{index-notation}{true}%
\newcommand{\ifindex}{\ifthenelse{\boolean{index-notation}}}
\newcommand{\Tref}{T_{\rm{L}}}
\newcommand{\ci}{{\bm{c}_i}}
\newcommand{\vi}{{\bm{v}_i}}
\newcommand{\x}{\bm{x}}
\newcommand{\e}{e}%
\newcommand{\ab}{\alpha\beta}
\newcommand{\dt}{\partial_t^{(1)}}
\newcommand{\dtt}{\partial_t^{(2)}}
\newcommand{\dalpha}{\partial_\alpha^{(1)}}
\newcommand{\dbeta}{\partial_\beta^{(1)}}
\newcommand{\dgamma}{\partial_\gamma^{(1)}}
\newcommand{\lp}{\left(}%
\newcommand{\rp}{\right)}%
\newcommand{\red}[1]{\textcolor{black}{#1}}

\Title{Kinetic Simulations of Compressible Non-Ideal Fluids: From Supercritical Flows to Phase-Change and Exotic Behavior}

\Author{Ehsan Reyhanian, Benedikt Dorschner and Ilya Karlin$^\ast$}

\AuthorNames{Ehsan Reyhanian, Benedikt Dorschner and Ilya karlin}

\address{Department of Mechanical and Process Engineering, ETH Zurich, 8092 Zurich, Switzerland}

\corres{Correspondence: ikarlin@ethz.ch}

\abstract{
\red{We investigate a kinetic model for compressible non-ideal fluids} [\url{DOI: 10.1103/PhysRevE.102.020103}]. 
\red{The model imposes the local thermodynamic pressure through a rescaling of the particle’s velocities,} 
which accounts for both long- and short-range effects and hence full thermodynamic consistency. 
The model is fully Galilean invariant and treats mass, momentum and energy as local conservation laws.
\red{The analysis and derivation of the hydrodynamic limit is followed by the assessment of 
accuracy and robustness through benchmark simulations ranging from Joule-Thompson effect to a phase-change and 
high-speed flows.} \red{In particular, we show the direct simulation of the inversion line of a van der Waals gas followed by simulations of phase-change such as the one-dimensional evaporation of a saturated liquid, nucleate and film boiling and eventually, we investigate the stability of a perturbed strong shock front in two different fluid mediums. In all of the cases, we find excellent agreement with the corresponding theoretical analysis and experimental correlations.}
We show that our model can operate in \red{the} entire phase diagram, including super- as well as sub-critical regimes and inherently captures 
phase-change phenomena.
}


\begin{document}
\section{Introduction}
The lattice Boltzmann method (LBM) is a kinetic-theory approach to the simulation
of hydrodynamic phenomena with applications ranging from turbulence \cite{Atif-turbulence,dorschner2016entropic} to microflows \cite{Kunert-micro,Harting-micro} and multiphase flows \cite{Sbragaglia-multiphase,Biferale-multiphase,Benzi-multiphase,Ali-multiphase}. The fully discretized kinetic equations evolve particle distribution functions (populations) $f_i(\x,t)$, which are associated with a set of discrete velocities $\bm{c}_i$, according to a simple stream-and-collide algorithm  and recover the Navier-Stokes equations in the hydrodynamic limit \cite{succi2018lattice}.

\red{While LBM has proven successful in a wide range of fluid mechanics problems
\cite{kruger2017lattice,succi2018lattice}, 
it is well-known that the fixed velocity set restricts conventional LB models to low-speed incompressible flows \cite{succi2018lattice}.
This promoted significant research efforts, which were directed towards the development of compressible LB models \cite{prasianakis2008lattice,frapolli2015entropic,pond,feng2019hybrid,wilde2020semi},
but they are typically limited to ideal gas. 
A genuine LB model, which can capture both compressible and non-ideal fluids has been lacking for long.}
\red{However, in many scientific and engineering applications the ideal-gas assumption is no longer valid and real-gas effects have to be taken into account.}
This includes phenomena such as rarefaction shock waves \cite{bates1999some,zhao2011admissible,guardone2002roe,zamfirescu2008admissibility}, acoustic emission instability \cite{bates-prl2000,bates2007instability}, inversion line (change of sign of the Joule-Thomson coefficient), phase transition, surface tension and super-critical flows. 

\red{While LB models for non-ideal gases have been subject to many studies in the literature, they are mostly restricted to incompressible flows. 
In the incompressible regime, two main approaches for non-ideal gases exist: pressure-based methods \cite{swift1995lattice,holdych1998improved} and forcing methods \cite{guo2002forcing,huang2011forcing,lycett2015improvedforcing,wagner2006investigation}}. 
\red{ Pressure-based methods were pioneered by Swift et al. \cite{swift1995lattice} and alter the equilibrium populations such that the full non-ideal pressure tensor, including the non-ideal equation of state (EOS) and the Korteweg stress, are recovered.}
\red{However, it was soon realized that these methods are not Galilean invariant \cite{holdych1998improved,inamuro2000galilean,wagner2006investigation} and lack thermodynamic consistency \cite{he2002thermodynamic}. 
Although various improvements have been proposed in the literature (see, e.g., \cite{swift1996lattice,holdych1998improved,inamuro2000galilean,wagner2006investigation}) their range of validity and stability remains limited \cite{wagner2006investigation}. }
On the other hand, forcing methods account for the deviation from the ideal-gas pressure by an appropriate (non-local) force term, which is introduced in the  kinetic equations. \red{Forcing methods are generally more stable than the pressure-based methods and the Galilean invariance error can be reduced effectively if augmented with appropriate correction terms \cite{wagner2006investigation}. 
Promising results have been obtained for forcing methods in various of applications}, ranging from droplet collisions at relatively large density ratios \cite{Ali-drop-collide,bosch2018entropic} to droplet impact on textured \cite{Ali-jfm} and flexible surfaces \cite{benedikt-fsi}.

\red{The aforementioned models have also been extended to thermal multiphase flows, including phase change. 
For instance, a common approach is to solve the temperature equation by conventional finite difference or finite volume schemes, which is then coupled to the flow field by a non-ideal EOS. 
As shown in \cite{zhang2003thermal,fei2020mesoscopic}, one can capture nucleate, transient and film boiling. 
Another common approach is to use a second set of population for the temperature equation, which is combined 
with additional source terms to account for phase change \cite{kamali2013thermal,safari2013extended,kupershtokh2018thermal}.
Under the low Mach conditions, these methods are commonly associated with simplifying assumptions such as neglecting the viscous heat dissipation \cite{kupershtokh2018thermal} or the pressure work \cite{safari2013extended,reyhanian2017investigation} which lead to a tailored form of the energy equation. 
}
\red{
Therefore, these models are not able to capture high-speed compressible flow of non-ideal fluids, where a complex temperature field with a wide range of values is expected to emerge.}

To mitigate these shortcomings, 
we recently proposed a novel method for non-ideal compressible fluid dynamics \cite{reyhanian2020thermokinetic} based
on adaptive discrete velocities in accordance to local flow conditions.
In contrast to the aforementioned schemes, the model features full Galilean-invariance and is thermodynamically consistent. %
As a consequence \red{of the model's construction}, the full energy-equation of a non-ideal fluid is recovered, which means that no additional phase-change model is required. 
This enables us to capture a large range of flow regimes, which we aim to explore in this paper.
\red{While basic validation was conducted in \cite{reyhanian2020thermokinetic} 
, we extend this analysis here 
and assess
the model's performance for super-critical flows, throttling, phase change and shock-stability.}

The paper is structured as follows: 
Section \ref{sec:Methodology} provides an in-depth analysis of the model. We start with a presentation of the discrete kinetic equations in Sec. \ref{subsec: Kinetic equations}, followed  by the Chapman-Enskog analysis and the derivation of the hydrodynamic limit in Sec. \ref{subsec:chapman}. Numerical benchmarks including the simulation of the Joule-Thomson effect, phase-change and high-speed flows are presented in Sec. \ref{sec:results}. 
Finally, conclusions are drawn in Sec. \ref{sec:conclusion}.
\section{Methodology}\label{sec:Methodology}
\subsection{Kinetic equations}\label{subsec: Kinetic equations}
Our thermokinetic model of non-ideal fluids \cite{reyhanian2020thermokinetic} is based on the so-called Particles-on-Demand (PonD) method \cite{pond},
which constructs the particle's velocities relative to the reference frame (gauge) $\lambda = \{T,\bm{u}\}$, where $T$ is the local temperature and $\bm{u}$ is the local flow velocity.
While the former leads to thermodynamic consistency, the latter guarantees Galilean invariance.
In addition, in the local reference frame, the local equilibrium becomes exact and solely dependent on the density. 
This is in contrast to classical LBM where one typically resorts to a truncated polynomial.
The populations can be transformed between different reference frames by requiring the moments to be independent of the reference frame \cite{pond}. 
In \cite{reyhanian2020thermokinetic}, we generalized this concept to encompass the thermodynamics of non-ideal fluids by defining the new set of discrete velocities as
\begin{equation} \label{eq particle velocity}
{\vi=\sqrt{\frac{p}{\rho \Tref}}\ci+{\bm{u}},}
\end{equation}
where $p(\bm{x},t)$ is the local thermodynamic pressure, $\rho(\bm{x},t)$ is the local density and $\Tref$ is a lattice reference temperature, a constant known for any set of speeds $\mathcal{C}=\{\ci, i=1,...,Q \}$, and $\bm{u}(\bm{x},t)$ is the local flow velocity.

A two-population approach is employed in this study. While $f$ populations maintain the density and the momentum field, the $g$ populations carry the total energy. 
As in classical LBM, a stream and collide algorithm is used to evolve the populations in time. 
In particular, we use a semi-Lagrangian approach for advection along the characteristics \cite{kramer2017semi,di2018simulation} at the monitoring point $(\bm{x},t)$, which reads
\begin{eqnarray}
    f_i ^\lambda (\x,t) &=& \widetilde{f_i} ^\lambda (\x-\vi \delta t,t-\delta t), \label{eq f streaming} \\
    g_i ^\lambda (\x,t) &=& \widetilde{g_i} ^\lambda(\x-\vi \delta t,t-\delta t), \label{eq g streaming}
\end{eqnarray}
while the Bhatnagar-Gross-Krook (BGK) model is employed for the collision step
\begin{eqnarray} \label{eq f collision}
f_i (\x,t) &=& f_i ^\lambda (\x,t) + \omega (f_i^{\rm eq} - f_i ^\lambda (\x,t)) + S_i^\lambda,\\
g_i(\bm{x},t) &=& g_i ^\lambda (\bm{x},t) + \omega (g_i^{\rm eq} - g_i ^\lambda (\bm{x},t)) + G_i ^\lambda {\delta}t \label{eq g collision},
\end{eqnarray}
where $\{f_i^{eq},g_i^{eq}\}$ denote the equilibrium populations. The source terms $\{S_i^\lambda$,$G_i^\lambda\}$ are used to account for the effect of surface tension in the momentum equation and a correction term in the energy equation, respectively. \red{Details will be provided in sections \ref{sec:heat_correction} and \ref{sec:surface_tension}}.

It is important to note that since the discrete velocities \eqref{eq particle velocity} depend on the local flow field (pressure, density and velocity), 
the departure point $\bm{x}_d =\bm{x}-\vi\delta t $ does not necessarily coincide with a grid node. 
Thus, the populations at the departure point need to be reconstructed and we use the general interpolation scheme 
\begin{align}
    \{ \widetilde{f}^\lambda (\bm{x}_d,t),\widetilde{g}^\lambda (\bm{x}_d,t)\}= \sum_{p=0}^{N} \Lambda(\bm{x}_d-\bm{x}_p)\mathcal{G}_{\lambda_p}^\lambda \{f^{\lambda_p} (\bm{x}_p,t),g^{\lambda_p} (\bm{x}_p,t)\},\label{eq: reconstruction}
\end{align}
where $\bm{x}_p$, $p=0,...,N$ denote the collocation points (grid points) and $\Lambda$ is the interpolation kernel. 
Notice that the population\red{s} at the collocation points are, in general, not in the same reference frame as the populations at the monitoring points. 
Thus, during the reconstruction step\red{,} populations are transformed from the reference \red{frame} $\lambda_p$ to $\lambda$ through the transformation 
matrix $\mathcal{G}_{\lambda_p}^\lambda$ \cite{pond}.
In general, a set of populations at gauge $\lambda$ can be transformed to another gauge $\lambda^\prime$ by matching \red{the} $Q$ linearly independent moments
\begin{align}
    \bm{M}_{mn}^\lambda = \sum_{i=1}^Q f_i^\lambda v_{ix}^m v_{iy}^n,
\end{align}
where $m$ and $n$ are integers. This may be written in the matrix product form as $\bm{M}^\lambda = \mathcal{M}_\lambda f^\lambda$ where $\mathcal{M}$ is the $Q\times Q$ linear map. 
Requiring that the moments must be independent from the choice of the reference frame, leads to the matching condition
\begin{align}
    \mathcal{M}_{\lambda^\prime}f^{\lambda^\prime} = \mathcal{M}_{\lambda}f^{\lambda}, 
\end{align}
which \red{yields} the transformed populations
\begin{align}
    f^{\lambda^\prime} = \mathcal{G}_\lambda^{\lambda^\prime}f^\lambda=\mathcal{M}_{\lambda^\prime}^{-1}\mathcal{M}_{\lambda}f^{\lambda}.
\end{align}
\red{Finally, we comment that the choice of the interpolation kernel is not the focus of the present study.} For simplicity, we use the third-order Lagrange polynomials in what follows, unless stated otherwise.\\

\begin{figure}
    \centering
     \includegraphics[width=1\linewidth]{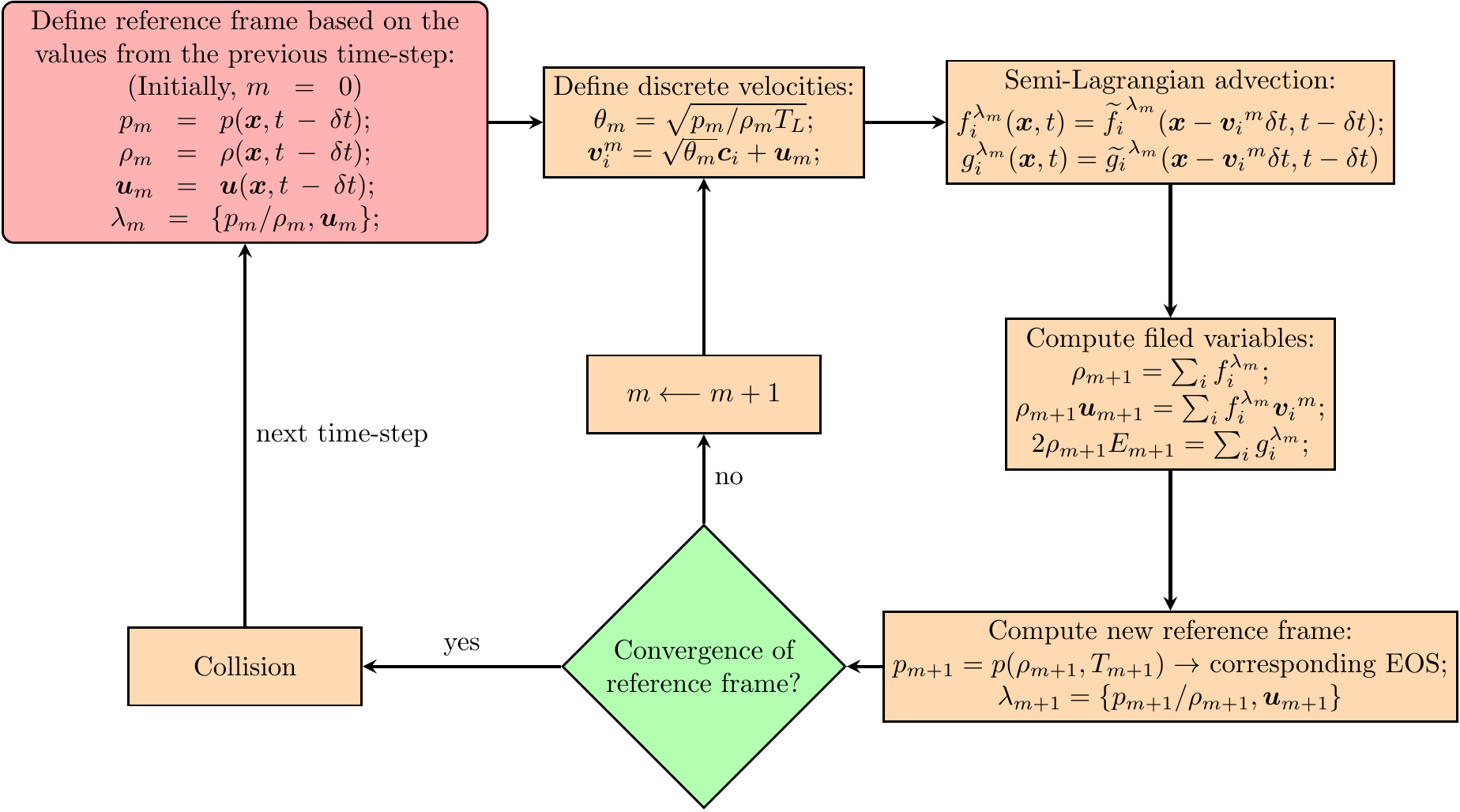}
    \caption{Flowchart of the semi-Lagrangian advection using the predictor-corrector algorithm.}
    \label{fig:flowchart}
\end{figure}

With the transformations defined, we are set for the advection scheme, where the local gauge is found iteratively using a 
predictor-corrector scheme. The full algorithm is depicted in Fig. \ref{fig:flowchart}.
Initially, the discrete velocities \eqref{eq particle velocity} are defined relative to the gauge $\lambda_0=\{p_0/ \rho_0, \bm{u}_0\}$ based on the pressure, density and velocity field from the previous time step. Once the discrete velocities $\vi^0$ are set, the semi-Lagrangian advections \eqref{eq f streaming} and \eqref{eq g streaming} are performed. 
With the new populations at the monitoring point, the density, momentum and total energy are evaluated by taking the corresponding moments of each population
\begin{eqnarray}
\rho_1 &=& \sum_{i=1}^{Q} f_i^{\lambda_0}, \\
\ifindex{\rho_1 u_{\alpha_1} &=& \sum_{i=1}^{Q} f_i^{\lambda_0} v_{i\alpha}^0 , \\}
{\rho_1 \bm{u}_1 &=& \sum_{i=1}^{Q} f_i \vi^0 , \\}
2\rho_1 e_1 + \rho_1 u_1^2 &=& \sum_{i=1}^{Q} g_i^{\lambda_0},
\end{eqnarray}
where $e = e(\rho,T)$ is the internal energy of a non-ideal fluid.
Subsequently, the pressure can be evaluated using the EOS of choice \red{with} the updated values for density and temperature. 
Finally, we can define the corrector gauge as $\lambda_1 = \{p_1/\rho_1,\bm{u}_1\}$ with $\bm{v}_i^1 = \sqrt{ p_1/(\rho_1 T_L)} \bm{c}_i +\bm{u_1}$.
This predictor-corrector step is iterated until the convergence of the gauge is achieved, where the limit gauge is the co-moving reference frame. Once the co-moving reference frame is determined, the advections \eqref{eq f streaming} and \eqref{eq g streaming} are completed.
Finally, the collision step in the co-moving reference frame follows. For the $f$ populations the local equilibrium takes the exact form
\begin{equation}
f_i^{\rm eq} = \rho W_i, \label{eq:feq}
\end{equation}
where $W_i$ are lattice weights, which are known for any velocity set.
The equilibrium for the $g$ population is derived using Grad's approximation \cite{shan1998discretization,ansumali2002kinetic,grad1949kinetic} 
for the new discrete velocities \eqref{eq particle velocity}. Thus, the equilibrium populations are constructed from the moments
\begin{align}
\ifindex
{
    M^{\rm eq} =\sum_{i=1}^Q g_i^{\rm eq}, \ q_{\alpha}^{\rm eq} =\sum_{i=1}^Q g_i^{\rm eq} v_{i\alpha},\  R_{\alpha\beta}^{\rm eq} =\sum_{i=1}^Q g_i^{\rm eq}v_{i\alpha}v_{i\beta},
}
{
    M^{\rm eq} =\sum_{i=0}^Q g_i^{\rm eq}, \ \bm{q}^{\rm eq} =\sum_{i=0}^Q g_i^{\rm eq} \vi,\  \bm{R}^{\rm eq} =\sum_{i=0}^Q g_i^{\rm eq}\vi\vi,
}
    \label{Appendix:moments}
\end{align}
where the explicit relations for the equilibrium moments are given by
\begin{align}
\ifindex
    {
M^{\rm eq} &= 2\rho E, \label{eq:total energy}\\
q_{\alpha}^{\rm eq}&= 2\rho u_\alpha H, \label{eq:heatflux}\\
    R_{\alpha\beta}^{\rm eq}&= 2\rho u_\alpha u_\beta \left( H+p/\rho\right) + 2pH\delta_{\alpha\beta}, \label{eq:highordermoment}
    }
    {
    M^{\rm eq} &= 2\rho E, \label{eq:total energy}\\
    \bm{q}^{\rm eq}&= 2\rho \bm{u} H, \label{eq:heatflux}\\
    \bm{R}^{\rm eq}&= 2\rho \bm{u}\bm{u} \left( H+p/\rho\right) + 2pH\bm{I}, \label{eq:highordermoment}
    }
\end{align}
\red{where $E = e + u^2/2$ is the total energy and $H = E + p/\rho$ is the total enthalpy.}
\ifindex{}{where $\bm{I}$ is the unit tensor.} Here, we shall define a second-order polynomial based on the general discrete velocities \ifindex{$v_{i\alpha} = \sqrt{\theta}c_{i\alpha} + u_\alpha$,}{$\vi = \sqrt{\theta}\ci + \bm{u}$,}
\begin{align}
\ifindex{
    g_i^{\rm eq} = W_i \left[ M^{\rm eq} + M_\alpha(v_{i\alpha}-u_\alpha) + M_{\alpha\beta}( v_{i\alpha} v_{i\beta} - \theta\Tref\delta_{\ab}-u_\alpha u_\beta) \right],
    }
    {
    g_i^{\rm eq} = W_i \left[ M^{\rm eq} + \bm{M}\cdot(\vi-\bm{u}) + \bm{N}:( \vi\vi - \theta\Tref\bm{I}-\bm{u}\bm{u}) \right],
    }
\label{Appendix:geq}
\end{align}
where it can easily be observed that the zeroth-order moment is already recovered. To satisfy the higher order moments, we must solve for \ifindex{$M_\alpha$ and $M_{\ab}$}{$\bm{N}$ and $\bm{M}$}. Substituting \eqref{Appendix:geq} in \eqref{Appendix:moments} gives
\begin{align}
\ifindex
   {
    q_\alpha^{\rm eq} &= M^{\rm eq} u_\alpha + M_\alpha \theta\Tref + 2\theta\Tref M_{\ab}u_\beta, \label{Appendix:q}\\
    R_{\ab}^{\rm eq}  &= M^{\rm eq} (\theta\Tref\delta_{\ab}-u_\alpha u_\beta) + q_\alpha^{\rm eq} u_\beta + q_\beta^{\rm eq} u_\alpha + 2\theta^2\Tref^2 M_{\ab}. \label{Appendix:R}
    }
    {
    \bm{q}^{\rm eq} &= M^{\rm eq} \bm{u} + \bm{M} \theta\Tref + 2\theta\Tref \bm{N}\cdot\bm{u}, \label{Appendix:q}\\
    \bm{R}^{\rm eq}  &= M^{\rm eq} (\theta\Tref\bm{I}-\bm{u}\bm{u}) + \bm{q}^{\rm eq} \bm{u} + \bm{u}\bm{q}^{\rm eq} + 2\theta^2\Tref^2 \bm{N}. \label{Appendix:R}
    }
\end{align}
Considering the explicit relations of the moments given by Eqs. \eqref{eq:total energy}-\eqref{eq:highordermoment} and taking into account the scaling $\theta = p / \rho\Tref$ one obtains
\begin{align}
\ifindex{
    M_\alpha &= 0, \nonumber\\
    M_{\ab} &= \rho \delta_{\ab}.
    }
    {
    \bm{M} &= 0, \nonumber\\
    \bm{N} &= \rho \bm{I}.
    }
\end{align}
Finally, upon substitution in Eq. \eqref{Appendix:geq},  $g_i^{\rm eq}$ is obtained as
\begin{align}
g_i^{\rm eq} = \rho W_i \left[2\e - {D}({p}/{\rho}) +  v_i^2\right], \label{eq:geq}
\end{align}
where $D$ is the space dimension.
As mentioned earlier, the source term $G_i$ in Eq. \eqref{eq g collision} is responsible for correction of the energy equation. 
\red{Here, we only derive the formulation and further details are discussed in section \ref{sec:heat_correction}. }
To derive an expression for the correction term in the co-moving reference frame, we follow the same method employed to express $g_i^{\rm eq}$. The pertinent moments of the correction term are
\begin{align}
\ifindex{
M_0 =\sum_{i=1}^Q G_i^\lambda, \ 0 =\sum_{i=1}^Q G_i^\lambda v_{i\alpha},\  0 =\sum_{i=1}^Q G_i^\lambda v_{i\alpha}v_{i\beta},}
{M_0 =\sum_{i=1}^Q G_i^\lambda, \ \bm{0} =\sum_{i=1}^Q G_i^\lambda \vi,\  \bm{0} =\sum_{i=1}^Q G_i^\lambda\vi\vi,}
\end{align}
Using relations \eqref{Appendix:q} and \eqref{Appendix:R}, one can write
\begin{align}
\ifindex
{
0 &= M_0 u_\alpha + M_\alpha \theta\Tref + 2\theta\Tref M_{\ab}u_\beta,\\
0 &= M_0 (\theta\Tref\delta_{\ab}-u_\alpha u_\beta) + 2\theta^2\Tref^2 M_{\ab},
}
    {
\bm{0} &= M_0 \bm{u} + \bm{M} \theta\Tref + 2\theta\Tref \bm{N}\cdot\bm{u},\\
\bm{0} &= M_0 (\theta\Tref\bm{I}-\bm{u}\bm{u}) + 2\theta^2\Tref^2 \bm{N},
}
\end{align}
which leads to the following solution
\begin{align}
\ifindex
{
    M_\alpha &= -M_0\frac{u^2}{\theta^2\Tref^2}u_\alpha, \nonumber\\
    M_{\ab} &= \frac{M_0}{2\theta^2\Tref^2}\left[ u_\alpha u_\beta - \theta\Tref\delta_{\ab} \right].
}
    {
\bm{M} &= -M_0\frac{u^2}{\theta^2\Tref^2}\bm{u}, \nonumber\\
    \bm{N} &= \frac{M_0}{2\theta^2\Tref^2}\left[ \bm{u}\bm{u} - \theta\Tref\bm{I} \right].    
}
\end{align}
Eventually, we can write the polynomial form of the correction term as
\begin{equation}
\ifindex{
    G_i^\lambda = {M_0}{W_i}{\lp 1+ \rho\frac{u_\alpha u_\beta c_{i\alpha}c_{i\beta}}{2pT_{\rm L}} -\frac{\rho v_i^2}{2p} +\frac{D}{2} \rp}, \label{eq g correction}}
{
    G_i^\lambda = {M_0}{W_i}{\lp 1+ \rho\frac{(\bm{u}\cdot\ci)^2}{2pT_{\rm L}} -\frac{\rho v_i^2}{2p} +\frac{D}{2} \rp}, \label{eq g correction}}
\end{equation}{}
where $M_0$ is the correction term in the energy equation.

We remind that the internal energy of a non-ideal fluid is now a function of both density as well as the temperature
\begin{equation}
    d\e=C_v dT + \left[T\lp\frac{\partial p}{\partial T}\rp_v - p\right]dv,
    \label{eq:real-gas internal energy diff}
    \end{equation}
where $C_v=({\partial e}/{\partial T})_v$ is the specific heat at constant volume and $v=1/\rho$ is the specific volume. 
For the sake of presentation, we shall at first neglect the interface energy and only consider $E = u^2/2 + \e$, where $E$ is the total energy, $\e=\e(s,v)$ is the local internal energy per unit of mass, $s$ is the entropy and \red{the} temperature is defined by $T=(\partial \e/\partial s)_v$.

In this paper, we use the classical van der Waals (vdW) EOS $ p=\rho RT/(1-b\rho) - a\rho^2$ to model non-ideal behavior of real-gases but others can be used analogously.
The constants are set to $a=2/49$, $b=2/21$ and $R=1$, where $a$ is the long-range attraction parameter,
$b$ represents the excluded-volume effect and $R$ is the specific gas constant. Considering the vdW EOS, we have
\begin{align}
    \left(\frac{\partial e_{vdw}}{\partial \rho}\right)_T &= T\lp\frac{\partial p}{\partial T}\rp_v - p =-a, \label{Appendix:e_vdw1}\\
    e_{vdw} &= \int C_v dT -a\rho, \label{Appendix:e_vdw2}
\end{align}
which suggests that $e_{vdw}=F(T)-a\rho$, where $F(T)$ is an arbitrary function of temperature. %
In other words, the internal energy of a vdW fluid is the sum of a density-dependent function and temperature-dependent function.

\subsection{Correction of the energy equation} \label{sec:heat_correction}
We first comment that the expression of heat flux recovered from the Chapman-Enskog analysis (see Section \ref{subsec:chapman}) without the correction term in the $g$ population is found as \ifindex{$q_\alpha^{\rm CE} = -\mu\partial_\alpha h$}{$\bm{q}^{\rm CE}=- \mu\nabla h$} where $\mu$ is the shear viscosity and $h=e+p/\rho$ is the specific enthalpy. In the limit of an ideal gas, this is equivalent to the Fourier law \ifindex{$q_\alpha^{\rm ig}=-k_{\rm ig}\partial_\alpha T$}{$\bm{q}_{\rm ig}=-k_{\rm ig}\nabla T $}, where $k_{\rm ig}=\mu C^{\rm ig}_p$ and hence the Prandtl number is fixed to ${\rm Pr}=\mu C_p^{\rm ig}/k_{\rm ig}=1$ due to the single relaxation time BGK collision model. However, considering the enthalpy of a real-gas as a function of pressure and temperature, we have \ifindex{$\partial_\alpha h =C_p\partial_\alpha T +v(1-\beta T)\partial_\alpha p$}{$\nabla h =C_p\nabla T +v(1-\beta T)\nabla p$}, where $\beta =v^{-1} \left(\partial v/\partial T\right)_p$ is the thermal expansion coefficient and $C_p = C_v + Tv \beta\lp{\partial p}/{\partial T}\rp_v $ is the specific heat at constant pressure. While one could eliminate the pressure part of the enthalpy by the correction term and only retain the temperature dependent part, it must be noted that the thermal expansion coefficient \textit{at} the critical point diverges, $\beta\to\infty$ and so does the specific heat $C_p\to \infty$. 
Hence, to recover the Fourier law, the post-collision of the $g$ population is augmented by the correction term $G_i^\lambda \delta t$, where $M_0 = \sum{G_i^\lambda}$ in Eq. \eqref{eq g correction} is set to \ifindex{$M_0=2\partial_\alpha(-\mu\partial_\alpha h + k \partial_\alpha T)$}{$= -2\nabla\cdot(\mu\nabla h) + 2\nabla\cdot (k\nabla T)$} and $k$ is the conductivity\red{, which is set independently.}

\subsection{Surface tension} \label{sec:surface_tension}
In order to describe two-phase flows in the sub-critical part of the phase diagram, the collision step for the $f$-populations (\ref{eq f collision}) is augmented with a source (forcing) term $S_i^\lambda$
\begin{align}
S_i^\lambda=\mathcal{G}_{\bm{u}+\delta\bm{u}}^{\bm{u}}[\rho W_i] -\rho W_i, \label{eq S-i}
\end{align}
where $\delta {\bm{u}} = \mathbf{F}/\rho \delta t$ is the change of the local flow velocity due to the force \ifindex{$F_\alpha = \partial_\beta K_{\ab}$}{$\mathbf{F}=\nabla \cdot \bm{K}$}, where
\begin{equation} \label{eq korteweg}
\ifindex
{K_{\ab} = \kappa\lp \Delta\rho+\frac{1}{2}|\nabla \rho|^2\rp\delta_{\ab}+\kappa\partial_\alpha\rho\partial_\beta\rho}
{\bm{K} =-\kappa(\rho \nabla\cdot\nabla \rho + \frac{1}{2} |\nabla \rho|^2)\bm{I} + \kappa\nabla\rho \otimes \nabla\rho,}
\end{equation}
is the Korteweg stress \cite{Ali-multiphase}, $\Delta=\nabla^2$ is the \red{Laplacian} and $\kappa$ is the surface tension coefficient. The first term on the R.H.S of equation (\ref{eq S-i}) denotes the transformation of equilibrium populations residing at the reference frame $"{\bm{u}}+\delta{\bm{u}}"$ to the reference frame $"{\bm{u}}"$ which \red{is equivalent to} the Exact Difference Method (EDM) \cite{kupershtokh2009equations}, adapted to the comoving reference frame. Having included the source term $S_i$, the actual fluid velocity is now shifted to $\hat{{\bm{u}}} = {\bm{u}} + \delta{\bm{u}}/2$\red{,} where ${\bm{u}}=1/\rho \sum f_i\vi$.\\
In the presence of \red{an} interface, the local equilibrium \eqref{eq:geq} is extended to account for the forcing $\bm{F}$. In order to do that, the same analogy used in the absence of the force term is employed here with the only difference that the velocity terms in the pertinent moments \eqref{eq:total energy}-\eqref{eq:highordermoment} are replaced by the modified velocities
\begin{align}
\ifindex{
    \hat{u}_\alpha = u_\alpha + \frac{F_\alpha\delta t}{2\rho}.
    }
{
    \hat{\bm{u}} = \bm{u} + \frac{\bm{F}\delta t}{2\rho},
    }
\end{align}
\red{In this setting,} the solution to relations \eqref{Appendix:q} and \eqref{Appendix:R} gives
\begin{align}
\ifindex{
    M_\alpha &= \frac{1}{(p/\rho)} \left[ F_\alpha(\hat{H}-u^2)\delta t - \left[ u_\alpha\delta t + \left(\hat{H}+\frac{p}{\rho}\right)\frac{\delta t ^2}{2p} F_\alpha \right] F_\beta u_\beta \right],\\
    M_{\ab} &= \rho\delta_{\ab} + \frac{\rho\delta t}{2p} \left[ u_\alpha F_\beta + F_\alpha u_\beta \right] + \frac{\rho\delta t ^2}{4p^2}F_\alpha F_\beta (\hat{H} + p/\rho),}
{
    \bm{M} &= \frac{1}{(p/\rho)} \left[ \bm{F}(\hat{H}-u^2)\delta t - \left[ \bm{u}\delta t + (\hat{H}+\frac{p}{\rho})\frac{\delta t ^2}{2p} \bm{F} \right] \bm{F}\cdot\bm{u} \right],\\
    \bm{N} &= \rho\bm{I} + \frac{\rho\delta t}{2p} \left[ \bm{u} \bm{F} + \bm{F} \bm{u} \right] + \frac{\rho\delta t ^2}{4p^2}\bm{F}\bm{F} (\hat{H} + p/\rho),}    
\end{align}
which can be written in a compact form using the expression \ifindex{$\hat{u}_\alpha F_\beta+\hat{u}_\beta F_\alpha = u_\alpha F_\beta + u_\beta F_\alpha+\delta t F_\alpha F_\beta / \rho$}{$\hat{\bm{u}}\bm{F}+\bm{F}\hat{\bm{u}} = \bm{u} \bm{F} + \bm{F}\bm{u} + \delta t \bm{F}\bm{F} / \rho$} and taking \ifindex{$G_{\ab} = \hat{u}_\alpha F_\beta+\hat{u}_\beta F_\alpha +\delta t F_\alpha F_\beta \hat{E}/2p$}{$\bm{G} = \hat{\bm{u}}\bm{F}+\bm{F}\hat{\bm{u}} +\delta t \bm{F}\bm{F} \hat{E}/2p$},
\begin{align}
\ifindex{
    M_\alpha &= \frac{\delta t}{(p/\rho)} \left[ F_\alpha \hat{H} - G_{\ab}u_\beta \right],\\
    M_{\ab} &= \frac{1}{(p/\rho)}\left[p\delta_{\ab} + \frac{\delta t}{2} G_{\ab}\right],
    }
{
\bm{M} &= \frac{\delta t}{(p/\rho)} \left[ \bm{F} \hat{H} - \bm{G}\cdot\bm{u} \right],\\
    \bm{N} &= \frac{1}{(p/\rho)}\left[p\bm{I} + \frac{\delta t}{2} \bm{G}\right],
}    
\end{align}
where $\hat{H}=\hat{E}+p/\rho = h+\hat{u}^2/2$. Finally, the extended equilibrium takes the form,
\begin{equation}
\ifindex{
g_i^{eq}=W_i \left[ 2\rho \hat{E}+ M_\alpha (v_{i\alpha}-u_\alpha) + M_{\ab}\left(v_{i\alpha}v_{i\beta}-\frac{p}{\rho}\delta_{\ab}-u_\alpha u_\beta\right) \right],
}
    {
 g_i^{eq}=W_i \left[ 2\rho \hat{E}+ \bm{M} \cdot (\vi-{\bm{u}}) + \bm{N}:\left(\vi\vi-\frac{p}{\rho}\bm{I}-{\bm{u}}{\bm{u}}\right) \right],
}
    \label{eq new geq}
\end{equation}
where
$\rho\hat{E}=\rho{E}+\hat{{u}}^2/2$ is based on the actual velocity of the flow. Note that in the absence of the force, the equilibrium (\ref{eq new geq}) simplifies to Eq. (\ref{eq:geq}). Consequently, the corresponding work of the added force is taken into account in the energy equation by modifying the correction term \eqref{eq g correction} with \ifindex{$M_0=2\partial_\alpha(-\mu\partial_\alpha h + k \partial_\alpha T) + 2\hat{u}_\alpha\partial_\beta K_{\ab}$}{$M_0 =2\nabla\cdot(-\mu\nabla h+k\nabla T)+2\hat{{\bm{u}}} \cdot \nabla \cdot \bm{K}$} . Finally, with the above modifications, the hydrodynamic equations for a two-phase system are recovered in \red{their} correct form. The evolution equations \eqref{eq:density}-\eqref{eq:temperature} together with the stress tensor \eqref{eq:stress tensor} remain intact \red{but all}  velocity terms $\bm{u}$ are replaced by the actual velocity $\hat{\bm{u}}$. Furthermore, the standard form of the total-energy conservation for a two-phase system \cite{liu2018thermal} is recovered
\begin{equation}
\ifindex
{\partial_t\left(\rho \hat{\mathcal{E}}\right)+\partial_\alpha\lp\rho \hat{\mathcal{E}}\hat{u}_\alpha+p\hat{u}_\alpha+\hat{\tau}_{\ab}\hat{u}_\beta+q_\alpha
          + K_{\ab}\hat{u}_\beta+\kappa\rho\partial_\beta\hat{u}_\beta\partial_\alpha \rho\rp=0,}
{\partial_t\left(\rho \hat{\mathcal{E}}\right)+\nabla\cdot\lp\rho \hat{\mathcal{E}}\hat{{\bm{u}}}+p\hat{{\bm{u}}}+\hat{\bm{\tau}}\cdot\hat{{\bm{u}}}+\bm{q}%
          + \bm{K}\cdot\hat{{\bm{u}}}+\kappa\rho\nabla\cdot\hat{{\bm{u}}}\nabla\rho\rp=0,}
\end{equation}
where $\rho \hat{\mathcal{E}}=\rho\hat{E}+ \frac{\kappa}{2} |\nabla\rho|^2$ accounts for the excess energy of the interface.

It is essential to mention that the van der Waals formulation of a real-gas \red{can yield} negative values of pressure for a range of temperatures in the subcritical region ($T_r<0.84375$) \red{and} a constant base-pressure must be added in order to have a meaningful evaluation of discrete velocities \eqref{eq particle velocity}. \red{Thus,} we redefine the pressure as \red{$p=p_{vdw}+\Bar{p}$ and choose $\Bar{p}$ such that  $p$ remains positive}. This will contribute to the internal energy according to relation \eqref{eq:real-gas internal energy diff} and the pressure-dependent part is re-evaluated as
\begin{align}
    T\lp\frac{\partial p}{\partial T}\rp_v - p = \frac{a}{v^2} - \Bar{p}.
\end{align}
Finally, the internal energy of a vdW fluid with base-pressure $\Bar{p}$ and constant specific heat $C_v$ \red{is given by}
\begin{align}
    \e = C_v T - a\rho - \frac{\Bar{p}}{\rho}.
\end{align}
However, the enthalpy of such a fluid remains intact since
\begin{align}
    h &= \e + \frac{p_{vdw}+\Bar{p}}{\rho},\\
    h &= C_v T - a\rho + \frac{p_{vdw}}{\rho}- \frac{\Bar{p}}{\rho}+\frac{\Bar{p}}{\rho},
    \end{align}
where the effect of the base-pressure is cancelled out in the evaluation of the enthalpy.

\section{Chapman-Enskog analysis}\label{subsec:chapman}
\unskip
\subsection{Excluding the forcing term}
Here we aim at \red{deriving} the macroscopic Navier-Stokes equations from the kinetic equations (\ref{eq f streaming}-\ref{eq g collision}). To this end, the pertinent equilibrium moments of \textit{f} and \textit{g} populations are required, which are computed as follows:
\begin{align}
    P_{\alpha\beta}^{eq}&=\sum_{i=1}^Q f_i^{eq} v_{i\alpha}v_{i\beta}= \rho u_\alpha u_\beta + p\delta_{\alpha\beta},\\
    Q_{\alpha\beta\gamma}^{eq}&=\sum_{i=1}^Q f_i^{eq} v_{i\alpha}v_{i\beta}v_{i\gamma}=\rho u_\alpha u_\beta u_\gamma + p[u\delta]_{\alpha\beta\gamma},\\
    q_{\alpha}^{eq}&=\sum_{i=1}^Q g_i^{eq} v_{i\alpha}= 2\rho u_\alpha H, \label{eq SM heatflux}\\
    R_{\alpha\beta}^{eq}&=\sum_{i=1}^Q g_i^{eq} v_{i\alpha}v_{i\beta}= 2\rho u_\alpha u_\beta \left( H+p/\rho\right) + 2pH\delta_{\alpha\beta}, \label{eq SM highordermoment}
\end{align}
where $[u\delta]_{\alpha\beta\gamma}=u_\alpha \delta_{\beta\gamma}+u_\beta \delta_{\alpha\gamma}+u_\gamma \delta_{\alpha\beta}$ and $H$ is the total enthalpy. First, we introduce the following expansions:
\begin{align}
    &f_i = f_i^{{(0)}}+ \epsilon f_i^{(1)} + \epsilon^2 f_i^{(2)}, \label{eq SM f expansion} \\
    &g_i = g_i^{(0)}+ \epsilon g_i^{(1)} + \epsilon^2 g_i^{(2)}, \label{eq SM g expansion}\\
    &\partial_t = \epsilon \dt + \epsilon^2 \dtt,\label{eq SM time expansion}\\
    &\partial_\alpha = \epsilon\dalpha.\label{eq SM spatial expansion}
\end{align}
Applying the Taylor expansion up to second order and separating the orders of $\epsilon$ results in
\begin{align}
    &\{f_i^{(0)},g_i^{(0)}\}=\{f_i^{eq},g_i^{eq}\},\\
    &\dt \{f_i^{(0)},g_i^{(0)}\} + v_{i\alpha}\dalpha \{f_i^{(0)},g_i^{(0)}\} = -(\omega/\delta t) \{f_i^{(1)},g_i^{(1)}\}, \label{eq SM chapman firstorder}\\
    &\dtt \{f_i^{(0)},g_i^{(0)}\} + \lp \dt+v_{i\alpha}\dalpha \rp \left(1-\frac{\omega}{2}\right)\{f_i^{(1)},g_i^{(1)}\} 
    =-(\omega/\delta t) \{f_i^{(2)},g_i^{(2)}\}.\label{eq SM chapman second-order}
\end{align}
The local conservation of density, momentum and energy imply
\begin{align}
\sum_{i=1}^Q\{ f_i^{(n)},g_i^{(n)}\}&=0, n\geq 1 \label{eq SM condition1},\\
\sum_{i=1}^Q f_i^{(n)}v_{i\alpha}&=0, n\geq 1. \label{eq SM condition2}     
\end{align}
Applying conditions (\ref{eq SM condition1}) and (\ref{eq SM condition2}) on equation (\ref{eq SM chapman firstorder}), we derive the following first order equations
\begin{align}
    &D_t^{(1)}\rho = -\rho\dalpha u_\alpha, \label{eq SM density-firstorder} \\
    &D_t^{(1)}u_\alpha = -\frac{1}{\rho}\dalpha p, \label{eq SM momentim-firstorder}\\
    &D_t^{(1)}T = -\frac{T}{\rho C_v} \lp \frac{\partial p}{\partial T} \rp_\rho \dalpha u_\alpha, \label{eq SM temperature-firstorder}
\end{align}
where $D_t^{(1)}=\dt+u_\alpha \dalpha$ is the first order total derivative. Subsequently, we can derive a similar equation for pressure considering that $p=p(\rho,T)$. This yields
\begin{equation}
    D_t^{(1)}p = \lp \frac{\partial p}{\partial \rho} \rp_T D_t^{(1)}\rho + 
                 \lp \frac{\partial p}{\partial T} \rp_\rho D_t^{(1)}T  
                 = -\rho \varsigma^2 \dalpha u_\alpha, \label{eq SM pressure-firstorder}
\end{equation}
where $\varsigma =\sqrt{\lp\frac{\partial p}{\partial \rho}\rp_s}$ is the speed of sound given by
\begin{equation} \label{eq SM speed of sound formula}
\varsigma =\sqrt{ \lp\frac{\partial p}{\partial \rho}\rp_T + \frac{T}{\rho^2 c_v}\lp\frac{\partial p}{\partial T}\rp_\rho ^2 }.
\end{equation}
The second order relations are obtained by applying the conditions (\ref{eq SM condition1}) and (\ref{eq SM condition2}) to equation (\ref{eq SM chapman second-order}),
\begin{align}
    &\dtt \rho = 0, \label{eq SM density-second-order} \\
    &\dtt u_\alpha = \frac{1}{\rho} \dbeta \left[ \delta t\lp\frac{1}{\omega}-\frac{1}{2}\rp \lp \dt P_{\alpha\beta}^{eq} + \dgamma Q_{\alpha\beta\gamma}^{eq} \rp\right], \label{eq SM momentum-second-order}\\
    &\dtt T = \frac{1}{2\rho C_v} \Bigg\{ \dalpha \left[ \delta t\lp\frac{1}{\omega}-\frac{1}{2}\rp \lp \dt q_{\alpha}^{eq} + \dbeta R_{\alpha\beta}^{eq} \rp\right] 
    -2\rho u_\alpha \dtt u_\alpha \Bigg\}. \label{eq SM temperature-second-order}    
\end{align}
Equations (\ref{eq SM density-firstorder}) and (\ref{eq SM density-second-order}) constitute the continuity equation. The non-equilibrium pressure tensor and heat flux in the R.H.S of equations (\ref{eq SM momentum-second-order}) and (\ref{eq SM temperature-second-order}) are evaluated using equations (\ref{eq SM density-firstorder}-\ref{eq SM pressure-firstorder}),
\begin{equation}
           \dt P_{\alpha\beta}^{eq} + \dgamma Q_{\alpha\beta\gamma}^{eq} =p\left(\dbeta u_\alpha
   +\dalpha u_\beta\right)
   +\left(p-\rho \varsigma^2\right)\dgamma u_\gamma \delta_{\alpha\beta},
\end{equation}
\begin{equation}
           \dt q_{\alpha}^{eq} + \dbeta R_{\alpha\beta}^{eq}=2\left(p-\rho \varsigma^2\right)\dgamma u_\gamma u_\alpha
           + 2p u_\beta\left(\dbeta u_\alpha+\dalpha u_\beta\right)+2p\dalpha h.
\end{equation}
Finally, summing up the contributions of density, momentum and temperature at the $\epsilon$ and $\epsilon^2$ orders and taking into account the correction to the energy equation \eqref{eq g correction}, we get the hydrodynamic limit of the model, which reads
	\begin{align}
	\ifindex{
&	D_t\rho=-\rho\partial_\alpha u_\alpha, \label{eq:density}\\
&	\rho D_t u_\alpha = -\partial_\alpha p -\partial_\beta \tau_{\ab},\label{eq:momentum} \\
&\rho C_v D_t T=-\tau_{\ab} \partial_\alpha u_\beta- T \left(\frac{\partial p}{\partial T}\right)_v \partial_\alpha u_\alpha
-\partial_\alpha q_\alpha,\label{eq:temperature}
	}
{
&	D_t\rho=-\rho\nabla\cdot{\bm{u}}, \label{eq:density}\\
&	\rho D_t{\bm{u}} = -\nabla p -\nabla\cdot\bm{\tau},\label{eq:momentum} \\
&\rho C_v D_t T=-\bm{\tau}:\nabla{\bm{u}}- T \left(\frac{\partial p}{\partial T}\right)_v \nabla\cdot{\bm{u}} -\nabla\cdot \bm{q},\label{eq:temperature}
}
\end{align}
where \ifindex{$D_t={\partial}_t+u_\alpha \partial_\alpha$}{$D_t={\partial}_t+{\bm{u}}\cdot\nabla$} is the material derivative, \ifindex{$q_\alpha=-k \partial_\alpha T$}{$\bm{q}=-k\bm{\nabla }T$} is the heat flux and the nonequilibrium stress tensor reads
	\begin{align}
\ifindex{
\tau_{\ab}= -\mu \lp\partial_\alpha u_\beta +\partial_\beta u_\alpha - \frac{2}{D}(\partial_\gamma u_\gamma)\delta_{\ab}\rp-\eta(\partial_\gamma u_\gamma)\delta_{\ab}. \label{eq:stress tensor}
}
    {
\bm{\tau}= -\mu \lp \nabla{\bm{u}}+\nabla{\bm{u}}^{\dagger}-\frac{2}{D}(\nabla\cdot\bm{u})\bm{I}\rp-\eta(\nabla\cdot{\bm{u}})\bm{I}. \label{eq:stress tensor}
}
	\end{align} 
The shear and bulk viscosity are
	\begin{align}
	\mu &= \lp \frac{1}{\omega}-\frac{1}{2} \rp p\delta t , \\
	\eta &= \lp \frac{1}{\omega}-\frac{1}{2} \rp \lp \frac{D+2}{D}-\frac{\rho \varsigma^2}{p} \rp p\delta t,
	\end{align}
respectively and $\varsigma=\sqrt{(\partial p/\partial \rho)_s}$ is the speed of sound. 
As expected, the bulk viscosity vanishes in the limit of ideal monatomic gas. Furthermore, one can observe that only the excluded volume part of the pressure $T(\partial p/ \partial T)_v$ contributes to the temperature equation \eqref{eq:temperature}, as expected.\\
\subsection{Including the forcing term}
As mentioned before, the force terms in the kinetic equations represent the interface dynamics. First, we recast the post-collision state of the \textit{f} population in the following form \cite{revisedchapman}
\begin{align}
&f_i^* (\x,t) = f_i (\x,t) + \omega (f_i^{eq}(\rho,\hat{\bm{u}}) - f_i (\x,t)) +\hat{S_i},\\
&\bm{\hat{u}}=\bm{u}+\frac{\bm{F}\delta t}{2\rho},\\
&\hat{S}_i=S_i-\omega\left(f_i^{eq}(\rho,\hat{\bm{u}})-\rho W_i\right),
\end{align}
where $[S_i=f_i^{eq}(\rho,{\bm{u}}+\bm{F}\delta t/\rho)-\rho W_i]$ and \ifindex{$F_\alpha = -\partial_\beta K_{\ab}$}{$\bm{F}=-\nabla\cdot\bm{K}$}. Here we expand the forcing term $\hat{S}^{(1)}_i=\epsilon\hat{S}^{(1)}_i$ in addition to the expansions (\ref{eq SM f expansion},\ref{eq SM time expansion},\ref{eq SM spatial expansion}). Similarly, we get the following relations at the orders of $\epsilon^0,\epsilon^1,\epsilon^2$, respectively
\begin{align}
    &f_i^{(0)}=f_i^{eq}(\rho,\hat{{\bm{u}}}),\\
    &\dt f_i^{(0)} + v_{i\alpha}\dalpha f_i^{(0)} = -(\omega/\delta t) f_i^{(1)}+\frac{1}{\delta t}\hat{S_i}^{(1)},\label{eq SM chapman first-order force}\\
    &\dtt f_i^{(0)} + \lp \dt+v_{i\alpha}\dalpha \rp (1-\frac{\omega}{2})f_i^{(1)}%
    +\frac{1}{2}\lp \dt+v_{i\alpha}\dalpha\rp \hat{S}_i^{(1)}
    =-(\omega/\delta t) f_i^{(2)}.
    \label{eq SM chapman second-order force}
\end{align}
It is important here to assess the solvability conditions imposed by the local conservations. Considering the moment-invariant property of the transfer matrix between the two gauges $\lambda=\{p/\rho,\bm{u}\}$ and $\hat{\lambda}=\{p/\rho,\hat{\bm{u}}\}$, one can easily compute
\begin{align}
    &\sum_{i=1}^Q f_i^{(0)}=\sum_{i=0}^Q f_i^{eq}(\rho,\hat{{\bm{u}}})=\rho,\\
    &\sum_{i=1}^Q f_i^{(0)} v_{i\alpha}=\sum_{i=0}^Q f_i^{eq}(\rho,\hat{{\bm{u}}})v_{i\alpha}=\rho\hat{u}_\alpha.
\end{align}
This implies that
\begin{align}
    \sum_{i=1}^Q f_i^{(n)}=0, n\geq 1, \label{eq SM condition_force 1} 
\end{align}
\begin{equation}
    \sum_{i=1}^Q f_i^{(n)}v_{i\alpha}= \Bigg\{
    \begin{tabular}{cc}
    $-\frac{\delta t}{2} F_\alpha^{(1)}$, & n=1, \\
    0, &  n$>$1.
    \end{tabular}
    \label{eq SM condition_force 2}
\end{equation}
According to the definition of $S_i$, the following moments can be computed:
\begin{align}
    \sum_{i=1}^Q \hat{S}_i^{(1)} &=0,\\
    \sum_{i=1}^Q \hat{S}_i^{(1)}v_{i\alpha} &=\delta t (1-\frac{\omega}{2})F_\alpha^{(1)},\\
    \sum_{i=1}^Q \hat{S}_i^{(1)}v_{i\alpha}v_{i\beta} &= \delta t(1-\frac{\omega}{2}) \lp \hat{u}_\alpha F_\beta +\hat{u}_\beta F_\alpha \rp+\frac{\omega \delta t^2F_\alpha F_\beta}{4\rho}.
\end{align}
Similarly, the first order equations of density and momentum are derived by applying the solvability conditions (\ref{eq SM condition_force 1}) and (\ref{eq SM condition_force 2}) on equations (\ref{eq SM chapman first-order force}) and (\ref{eq SM chapman second-order force}),
\begin{align}
    &\hat{D}_t^{(1)}\rho = -\rho\dalpha \hat{u}_\alpha, \label{eq SM density-firstorder-force} \\
    &\hat{D}_t^{(1)}\hat{u}_\alpha = -\frac{1}{\rho}\dalpha p+ \frac{1}{\rho}F^{(1)}_\alpha, \label{eq SM momentim-firstorder-force}
\end{align}
where $\hat{D}_t^{(1)}=\dt+\hat{u}_\alpha \dalpha$. At this point it is necessary to mention that since there is a force added to the momentum equation (in this case the divergence of the Korteweg stress), it should also be considered in the energy equation as well. Hence as mentioned before, $M_0$ in Eq. (\ref{eq g correction}) is modified to
\begin{align}
\ifindex
{M_0=2\partial_\alpha(-\mu\partial_\alpha h + k \partial_\alpha T) + 2\hat{u}_\alpha F_\alpha.}
{M_0 = -2\nabla\cdot(\mu\nabla h) + 2\nabla\cdot (k\nabla T)+2\hat{u}_\alpha F_\alpha.}
\end{align}
The equilibrium moments are modified as
\begin{align}
    &\sum_{i=1}^Q g_i^{eq}=2\rho \hat{E},\\
    q_{\alpha}^{eq}=&\sum_{i=1}^Q g_i^{eq} v_{i\alpha}= 2\rho \hat{u}_\alpha \hat{H}, \label{eq SM heatflux-force}\\
    R_{\alpha\beta}^{eq}=&\sum_{i=1}^Q g_i^{eq} v_{i\alpha}v_{i\beta}= 2\rho \hat{u}_\alpha \hat{u}_\beta \left( \hat{H}+p/\rho\right) + 2p\hat{H}\delta_{\alpha\beta}, \label{eq SM highordermoment-force}
\end{align}
where $\hat{E}=e+\hat{u}^2/2$ and $\hat{H}=\hat{E}+p/\rho$. With the changes mentioned so far, the first-order equation of temperature is derived as
\begin{align}
    \hat{D}_t^{(1)}T = -\frac{T}{\rho C_v} \lp \frac{\partial p}{\partial T} \rp_\rho \dalpha \hat{u}_\alpha. \label{eq SM temperature-firstorder-force}
\end{align}
Finally, in a similar manner \red{to} the case without the force, the macroscopic equations are recovered by collecting the equations of density, momentum and temperature at each order
	\begin{align}
	\ifindex{
	\hat{D}_t\rho&=-\rho\partial_\alpha \hat{u}_\alpha, \label{eq SM density-force}\\
	\rho \hat{D}_t \hat{u}_\alpha &= -\partial_\alpha p -\partial_\beta \hat{\tau}_{\ab}-\partial_\beta K_{\ab},\label{eq SM momentum-force} \\
\rho C_v \hat{D}_t T &=-\hat{\tau}_{\ab} \partial_\alpha \hat{u}_\beta- T \left(\frac{\partial p}{\partial T}\right)_v \partial_\alpha \hat{u}_\alpha
-\partial_\alpha q_\alpha,\label{eq SM temperature-force}\\
\hat{\tau}_{\ab} &= -\mu \lp\partial_\alpha \hat{u}_\beta +\partial_\beta \hat{u}_\alpha - \frac{2}{D}(\partial_\gamma \hat{u}_\gamma)\delta_{\ab}\rp-\eta(\partial_\gamma \hat{u}_\gamma)\delta_{\ab}.
	}
{
&\hat{D}_t\rho=-\rho\nabla\cdot\hat{{\bm{u}}}, \label{eq SM density-force}\\
    &\rho \hat{D}_t\hat{\bm{u}} = -\nabla p -\nabla\cdot\hat{\bm{\tau}} -\nabla\cdot \bm{K},\label{eq SM momentum-force} \\
    &\rho C_v \hat{D}_t T=-\hat{\bm{\tau}}:\nabla\hat{{\bm{u}}}- T \left(\frac{\partial p}{\partial T}\right)_v \nabla\cdot\hat{\bm{u}}-\nabla\cdot \bm{q},\label{eq SM temperature-force}\\
    &\hat{\bm{\tau}}= -\mu \lp \nabla\hat{{\bm{u}}}+\nabla\hat{{\bm{u}}}^{\dagger}-\frac{2}{D}(\nabla\cdot\hat{\bm{u}})\bm{I}\rp-\eta(\nabla\cdot\hat{{\bm{u}}})\bm{I}.
}
\end{align}
It should be noted that the error terms associated with the forcing are not shown here. For instance, as reported in the literature \cite{revisedchapman,lycett2014multiphase,wagner2006thermodynamic}, one can show that the error term in the momentum equation appears as $\nabla\cdot(\delta t^2\bm{F}\bm{F}/4\rho)$.

The total energy of the fluid is formulated by $\hat{\mathcal{E}}=e \lp T,v \rp+\hat{u}^2/2+E_\lambda$, where $E_\lambda=\kappa|\nabla\rho|^2/2$ is the non-local part corresponding to the excess energy of the interface. The evolution equation for the specific internal energy $e(T,v)$ can be obtained by considering equations (\ref{eq:real-gas internal energy diff},\ref{eq SM density-force},\ref{eq SM temperature-force})
\begin{equation}
\ifindex
{\rho \hat{D}_t e=-p\partial_\alpha\hat{u}_\alpha-\hat{\tau}_{\ab}\partial_\alpha \hat{u}_\beta-\partial_\alpha q_\alpha. \label{eq SM internal energy revolution}}
{\rho \hat{D}_t e=-p\nabla\cdot\hat{{\bm{u}}}-\hat{\bm{\tau}}:\nabla\hat{{\bm{u}}}-\nabla\cdot \bm{q}.
    \label{eq SM internal energy revolution}}
\end{equation}
From the momentum equation (\ref{eq SM momentum-force}) we get
\begin{equation}
\ifindex
{\frac{1}{2}\rho\hat{D}_t \hat{u}^2 = -\hat{u}_\alpha\partial_\alpha p -\hat{u}_\alpha\partial_\beta \hat{\tau}_{\ab}-\hat{u}_\alpha\partial_\beta K_{\ab},
    \label{eq SM kinetic energy revolution}}
{\frac{1}{2}\rho\hat{D}_t \hat{u}^2 = -\hat{{\bm{u}}}\cdot\nabla p -\hat{\bm{u}}\cdot\nabla\cdot\hat{\bm{\tau}}-\hat{\bm{u}}\cdot\nabla\cdot\bm{K},
    \label{eq SM kinetic energy revolution}}
\end{equation}
and the evolution of the excess energy can be computed using the continuity equation
\begin{equation}
\ifindex
{\rho\hat{D}_t E_\lambda=-K_{\ab}\partial_\beta\hat{u}_\alpha - \partial_\alpha \lp \kappa \rho \partial_\beta \hat{u}_\beta \partial_\alpha \rho \rp. \label{eq SM excess energy revolution}}
{\rho\hat{D}_t E_\lambda=-\bm{K}:\nabla\hat{{\bm{u}}}-\nabla\cdot\left(\kappa\rho\nabla\cdot\hat{{\bm{u}}}\nabla\rho\right). \label{eq SM excess energy revolution}}
\end{equation}
\red{Finally, upon summation of all three parts,} we get the full conservation equation for the total energy
\begin{align}
\ifindex
{\partial_t\left(\rho \hat{\mathcal{E}}\right)+\partial_\alpha\Big(\rho \hat{\mathcal{E}}\hat{u}_\alpha+p\hat{u}_\alpha+\hat{\tau}_{\ab}\hat{u}_\beta+
      K_{\ab}\hat{u}_\beta
      +\kappa\rho\partial_\beta\hat{u}_\beta\partial_\alpha\rho+ q_\alpha\Big)=0.}
{\partial_t\left(\rho \hat{\mathcal{E}}\right)+\nabla\cdot\Big(\rho \hat{\mathcal{E}}\hat{{\bm{u}}}+p\hat{{\bm{u}}}&+\hat{\bm{\tau}}\cdot\hat{{\bm{u}}}+
      \bm{K}\cdot\hat{{\bm{u}}}
      +\kappa\rho\nabla\cdot\hat{{\bm{u}}}\nabla\rho+\bm{q}\Big)=0.}
\end{align}
\section{Results and discussion}\label{sec:results}
In this section, we show validity of our model in a broad range of problems, which are chosen to probe
the correct thermodynamics as well as Galilean invariance:

\begin{itemize}
\item As a first test of basic thermodynamic consistency for non-ideal fluids, we simulate the inversion line of a vdW fluid, which is one of the classic thermodynamical concepts of non-ideal fluids. 
To capture this phenomenon it is crucial that the model recovers the correct energy equation and can operate in a  
wide range of pressures and temperatures in the super-critical part of the phase diagram.

\item Phase-change is the next fundamental process that is tested with our model.
It is important to remind that since the full energy equation is recovered by our kinetic equations, \red{phase-change} emerges naturally in the proposed scheme and no additional \emph{ad-hoc} phase-change model is required.
In addition, we probe fast dynamics with temperatures near the critical point, where phase-change happens on short time scales.

\item As a final test case we probe both thermodynamic consistency as well as Galilean invariance in supersonic flows. 
In particular, we study the behavior of a perturbed shock-front in both an ideal gas as well as a vdW fluid at Mach number 
Ma=3. In agreement with theory, our model shows to capture all regimes, including the exotic behaviors of a real fluid.
\end{itemize}
\red{For all simulations, we use the standard D2Q9 lattice, where $D=2$ denotes the spatial dimension and $Q=9$ is the number of discrete velocities.
}
\subsection{Inversion line}\label{subsec:inversion}
When a fluid passes through a throttling device, the value of the enthalpy remains constant in the absence of work and heat. 
During this so-called throttling process, the pressure of the fluid drops and the behavior of the temperature is characterized by the Joule--Thomson (JT) coefficient {$\mu = \left(\partial T / \partial P \right)_h $} \cite{cengel2007thermodynamics}. 
Depending on the sign and value of the JT coefficient, the temperature may increase, decrease or remain constant through the process. 
For ideal gas, the JT coefficient vanishes and thus the temperature does not change. 
On the other hand, for real gases, we need to  distinguish between three different regions in the $T-P$ diagram, corresponding to the different signs of the Joule--Thomson coefficient. 
Let us start by defining the inversion line as the locus of points where $\mu = 0$. Hence, crossing the inversion line  will
lead to a change of sign of $\mu$.
For the vdW EOS, one can derive the expression for the inversion line as 
\begin{equation}
    P_r = 24\sqrt{3T_r}-12T_r-27,
    \label{eq: inversion line}
\end{equation}
where the subscript "r" indicates that the quantities are reduced with respect to their values at the critical point. The critical values of pressure, temperature and density for a vdW fluid are $P_{cr} = a/27b^2, T_{cr}=8a/27Rb$ and $\rho_{cr}=1/3b$, respectively. In addition to the reduced variables, it is useful to define the non-dimensional enthalpy as
\begin{equation}
    \hat{h} = \frac{h}{RT_{cr}}= T_r \left[\frac{1}{\delta}+\frac{3}{3-\rho_r} \right]-\frac{9}{4}\rho_r,
\end{equation}
where $\delta=R/C_v$ is a constant. A closer assessment of (\ref{eq: inversion line}) reveals that the point with the maximum pressure on the inversion line corresponds to the following values, {$\left(P_r, \rho_r, T_r, \hat{h} \right) = \left(9.0, 1.0, 3.0, 11.25 \right)$}.

To test that our model captures these phenomena also numerically, we measure the value of the Joule--Thomson coefficient at different points in the $T-P$ diagram. 
We do this in two steps: in a first simulation, the flow is subjected to a positive acceleration under fixed density; hence the pressure drops and the quantity {$\left( \partial P / \partial T\right)_\rho $} is measured. In a second simulation, the isothermal speed of sound {$\left( \partial P / \partial \rho\right)_T $} is computed by introducing a small perturbation in the pressure field and measuring the velocity of the subsequent shock front. The Joule-Thomson coefficient is computed by
\begin{equation}
    \mu = -\frac{1}{C_p}\left[ \frac{1}{\rho}-T\frac{\left( \partial P / \partial T\right)_\rho}{\rho^2\left( \partial P / \partial \rho\right)_T} \right].
\end{equation}\label{eq: JT}
Finally, we use a simple Euler scheme to construct the isenthalpic lines with
\begin{equation}
    \Delta T \approx \mu \Delta P.
    \label{eq: Euler}
\end{equation}
The simulations are conducted for three different enthalpies, $\hat{h}=5,\hat{h}=11.25$ and $\hat{h}=15$. Figure \ref{fig:1} shows the measured values of the dimensionless Joule-Thomson coefficient at different reduced pressures up to the far supercritical value $P_r=15$. The comparison between the van der Waals theory and the simulation is excellent and thus
validates our scheme.
	\begin{figure}
		\centerline{\includegraphics[width=0.85\linewidth]{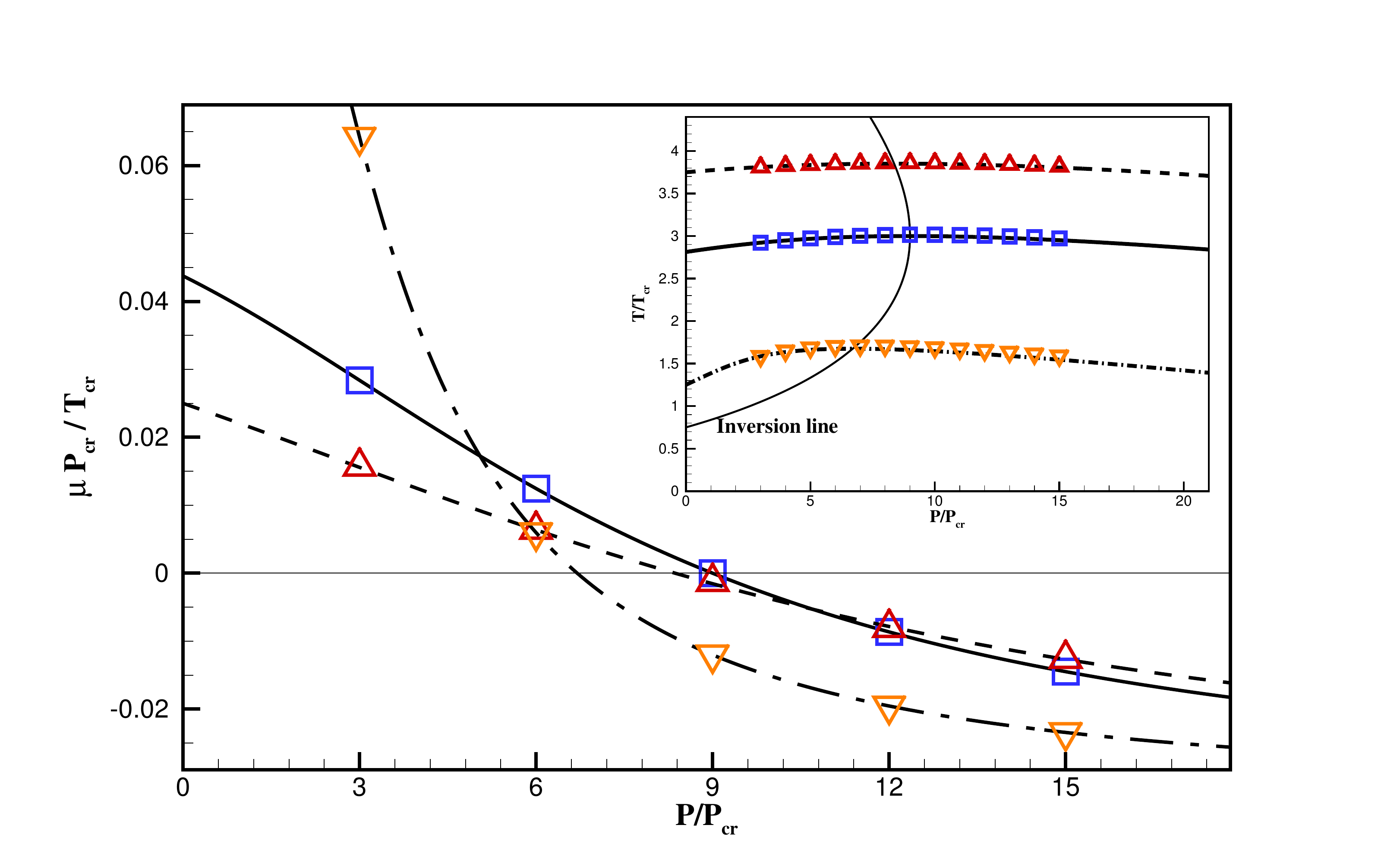}}
		\caption{Joule-Thomson coefficient against reduced pressure. The value of the Joule-Thomson coefficient along the dimensionless isenthalpic lines $\hat{h}=h/RT_{cr}$ was measured in a wide range of reduced pressures up to $P/P_{cr}=15$.
				Line: theory; Solid: $\hat{h}=11.25$; Dashed: $\hat{h}=15$; Dash dot: $\hat{h}=5$. Symbols: present method; Squares: $\hat{h}=11.25$; Triangles: $\hat{h}=15$; Inverted triangles: $\hat{h}=5$.
			Inset: Simulated lines of constant enthalpy on the $T_r-P_r$ (phase) diagram.}
		\label{fig:1}
	\end{figure}
\subsection{Phase change: one-dimensional Stefan problem}
In this section, we validate our model for phase-change problems, starting from the one-dimensional Stefan problem, 
where a liquid-vapor system is subjected to a heated wall on the vapor side.
The heat transfer from the wall leads to evaporation of liquid and the interface is moving away from the wall. The analytical solution for the liquid-vapor interface location with time is given by $x_i(t) = 2\beta\sqrt{\alpha_v t}$, where $\alpha_v$ is the diffusivity of the vapor and $\beta$ is the solution to \cite{welch2000volume}
\begin{equation}
    \beta \exp({\beta^2}) \text{erf}(\beta) = \frac{\text{St}}{\sqrt{\pi}},
    \label{eq: Stefan}
\end{equation}
where $\text{St} = C_{pv}\Delta T/h_{fg}$ is the Stefan number, $C_{pv}$ is the specific heat capacity of the vapor phase, $\Delta T$ is the temperature difference between the wall and the saturation temperature and $h_{fg}$ denotes the latent heat of evaporation. Simulations were carried out for three different Stefan numbers at fixed diffusivity. The choice of the Stefan number is directly related to the velocity of the interface:
\begin{align}
    u_i(t) = \frac{d}{dt}x_i(t) = \beta \sqrt{\frac{\alpha_v}{t}},
\end{align}
and hence the Mach number of the flow. 
Note that since our model is not restricted to low-speed flows,  we can accurately capture a wide range of Stefan numbers. Figure \ref{fig:Stefan} shows the location of the interface during evaporation compared to the analytical solution. The results of the simulation are in good agreement with the theory. We mention that the choice of parameters in our model such as the latent heat of evaporation $h_{fg}$ or the specific heat $C_p$ is not arbitrary and they are computed based on the thermodynamical state of the initial flow. For instance, the value of the Stefan number increases for a given temperature difference $\Delta T$ as we approach the critical point due to vanishing of $h_{fg}$ and diverging of $C_p$ at the critical point. 
\begin{figure}
  \centerline{\includegraphics[width=1
  \linewidth]{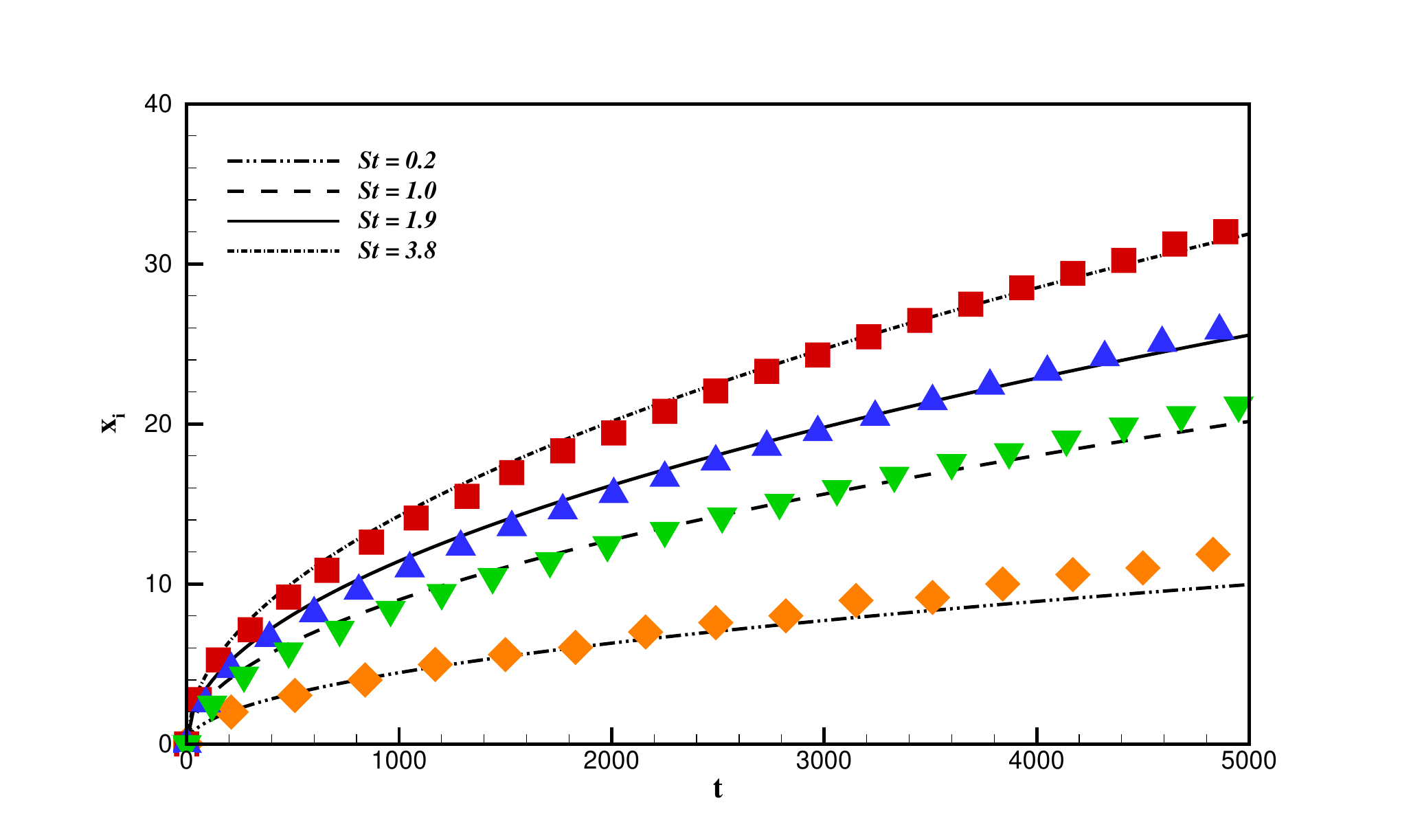}}
  \caption{Location of the interface versus time in lattice units for four different Stefan numbers. Line: analytical solution. Symbols: present method.}
\label{fig:Stefan}
\end{figure}
\subsection{Phase change: nucleate boiling}
Due to its importance in engineering  and real life applications, various boiling regimes have been the focus of many studies, both numerically and theoretically \cite{liu2018thermal,onuki2005dynamic,onuki2007dynamic}.
To validate our model, we choose a two-dimensional setup, where a liquid is in direct contact with a wall with high temperature \red{in} the middle of the wall. The schematic of the setup is shown in Fig. \ref{fig: Nucleating}a. The non-uniform temperature of the wall triggers a two-dimensional flow and the nucleus starts to appear and rise under the gravitational field. The nucleus continues to rise and grow until necking is achieved \red{and} the bubble is detaching from the nucleus. Once the first bubble is detached from the nucleus and released into the liquid, the nucleus continues to grow and releases a second bubble. This is a periodic process of bubble release, which is a function of surface tension, density ratio and the gravity. 
An empirical correlation for the bubble release frequency was found experimentally by Zuber \cite{zuber1963nucleate}
and reads
\begin{align}
    f^{-1} \approx \frac{d}{0.59} \left( \frac{\sigma g (\rho_l - \rho_v)}{\rho_l^2}  \right)^{-1/4},
\end{align}
where $d$ is the departure diameter and is itself proportional to $g^{-0.5}$ \cite{kocamustafaogullari1983pressure}, \red{ $\rho_l$ is the density of the liquid phase and $\rho_v$ is that of the vapor phase}. Hence, the bubble release period is proportional to $g^{-0.75}$. We consider \red{a} domain of $121\times 601$ points with time step $\delta t = 0.3$, conductivity $k=0.6$, specific heat $C_v=3$, viscosity $\nu = 0.005$, surface tension coefficient $\kappa = 0.0234$ and gravity $g = 0.0001$ in lattice units. The Jacob number is defined as
\begin{align}
    Ja = \frac{C_{pl}(T_w-T_{sat})}{h_{fg}},
\end{align}
where $C_{pl}$ is the specific heat of the liquid phase. The wall temperature is set to $T_w=1.5 T_{sat}$, where $T_{sat}$ is the saturation temperature and the initial temperature of the liquid is $T_{sat}=0.9T_{cr}$, which fixes the latent heat of evaporation. This choice of parameters leads to the Jacob number $Ja = 2.21$. Fig.\ \ref{fig: Nucleating}b illustrates a sequence of the bubble interface from the early times of the first nucleus development until the first bubble is released into the liquid. 
The bubble release period was measure for different values of gravity and the results are presented in 
in Fig.\ \ref{fig:boiling}. The comparison shows that the bubble release period is proportional to $g^{-0.75}$ in our numerical simulations which is in agreement
with the empirical~correlation.
\begin{figure}[h]
\begin{subfigure}[b]{0.5\textwidth}
    \centering
    \includegraphics[width=\linewidth]{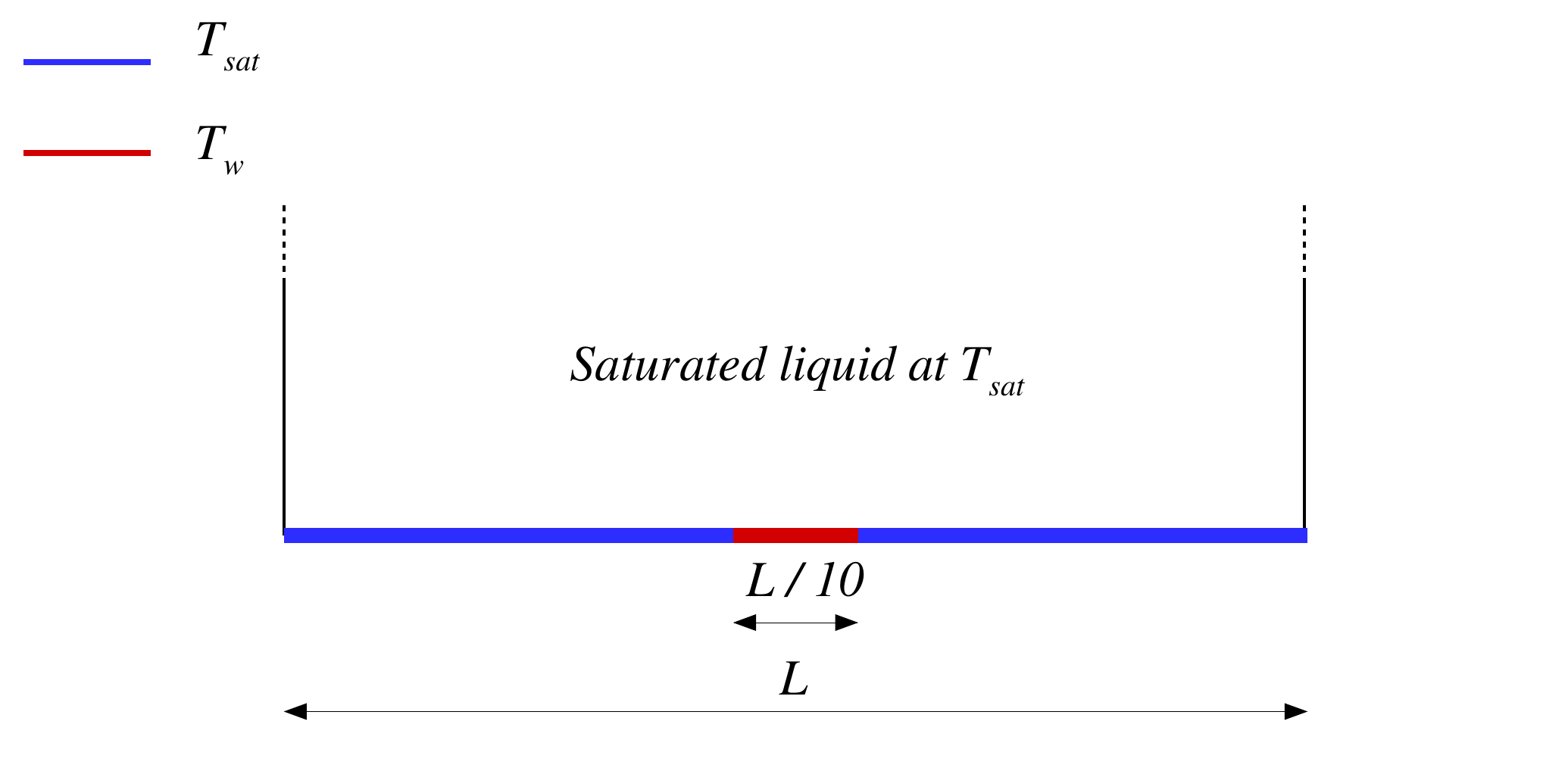}
    \caption{}
    \end{subfigure}
    \begin{subfigure}[b]{0.5\textwidth}
    \centering
    \includegraphics[clip, trim= 1.5cm 0.5cm 1.5cm 1.5cm, width=\linewidth]{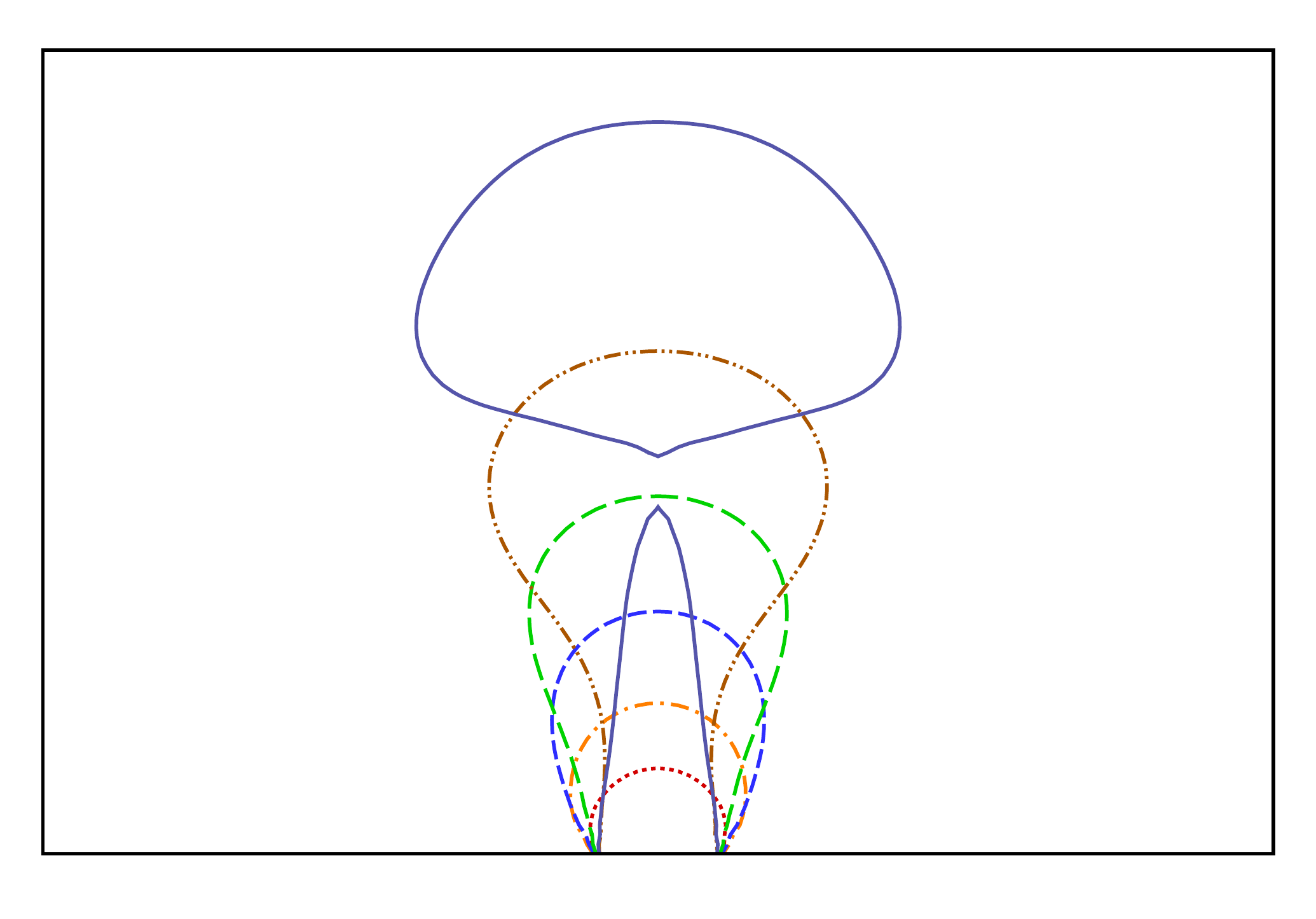}
    \caption{}
\end{subfigure}
\caption{(a) Schematic of the nucleate problem. (b) The interface of the vapor bubble during the nucleation, starting from the appearance of the first nucleus until the release of the first bubble. From bottom to top; Fine-dashed: time=600, Dash dot: time = 1200, Dashed: time=1800, Long-dashed: time=2400, Dash dot-dot: time=3000, Solid: time=3780. Times are measured in lattice units.}
\label{fig: Nucleating}
\end{figure}

\begin{figure}[H]
\centering
  \centerline{\includegraphics[clip, trim= 0.9cm 0.5cm 1.5cm 1.5cm, width=0.8\linewidth]{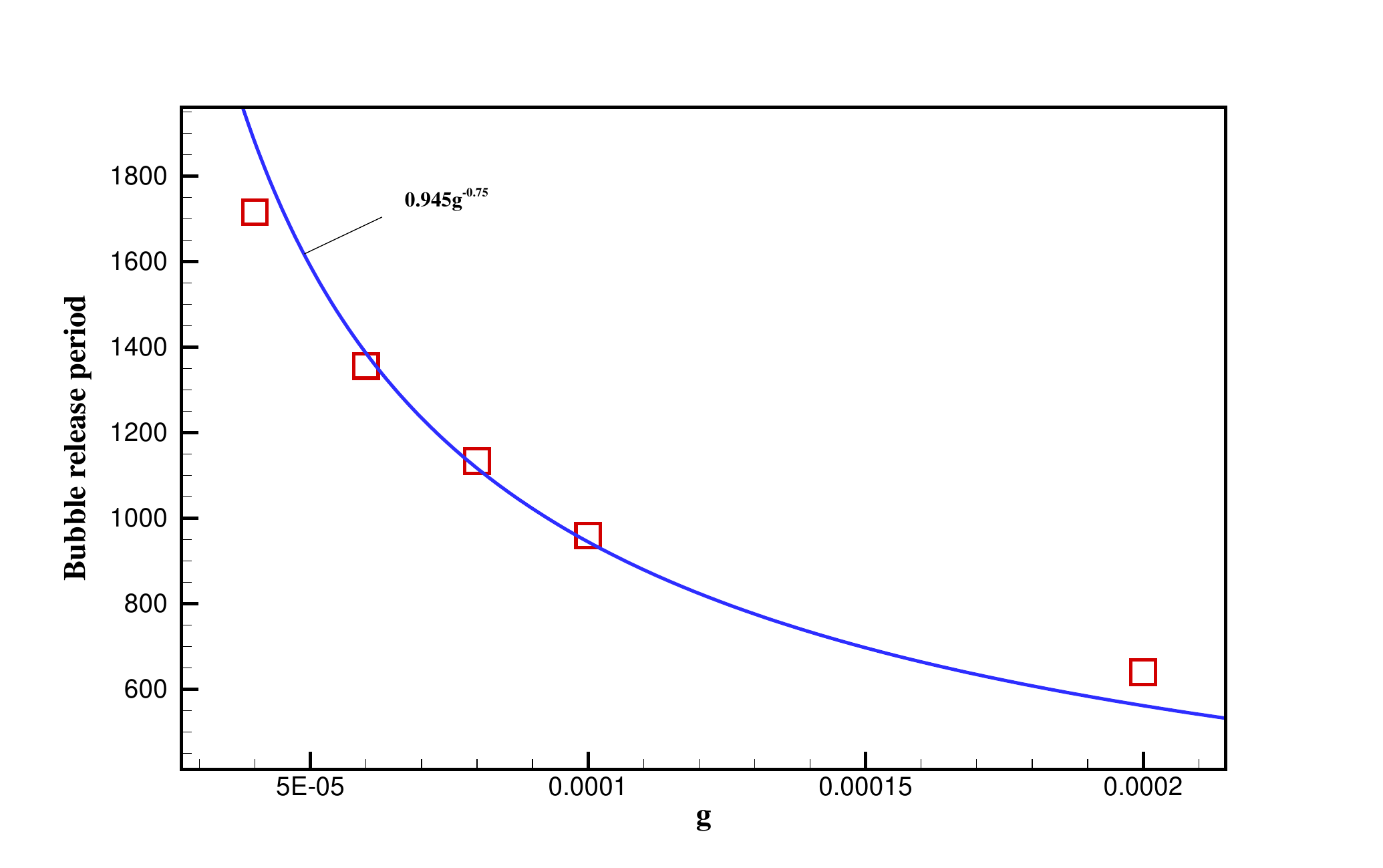}}
   \caption{Bubble release period against different gravity numbers. Symbols: Simulation. The solid line represents a function $0.945g^{-0.75}$.}
\label{fig:boiling}
\end{figure}
\subsection{Phase change: Single-mode film boiling}
As a final phase-change validation, we conduct simulations of film boiling, where a heated horizontal surface is covered by 
a thin layer of vapor. 
The liquid rests on top of the vapor and both phases are initially saturated. Phase-change then takes place at the liquid-vapor interface, where the heat is transported from the hot wall with temperature $T_w$, which is set to be above its saturation temperature $T_{sat}.$ The governing non-dimensional numbers are the Jacob, \red{the} Prandtl and the Grashof number \cite{esmaeeli2004computations}, 
\begin{align}
    Ja &= \frac{C_{pv}(T_w-T_{sat})}{h_{fg}},\\
    Pr &= \frac{\mu_v C_{pv}}{k_v},\\
    Gr &= \rho_v g (\rho_l - \rho_v)\frac{l_s^3}{\mu_v^2},
\end{align}
\red{which are defined for the vapor phase. The non-dimensional capillary length $l_s$ is defined as}
\begin{align}
    l_s = \sqrt{\frac{\sigma}{(\rho_l-\rho_v)g}},
\end{align}
and $t^* = t/\sqrt{l_s/g}$ is the dimensionless time. The well-known Klimenko correlation as proposed in  \cite{klimenko1982film} has the following form in the laminar flow regime ($Gr\le 4.03\times 10^5$)
\begin{align}
     &Nu_{\red{k}} = 0.1691 \lp \frac{Gr Pr}{Ja}\rp ^{1/3},  &Ja < 0.71, \label{eq:klimenko1}\\
     &Nu_{\red{k}} = 0.19 \lp Gr Pr\rp ^{1/3},   &Ja \geq 0.71,\label{eq:klimenko2}
\end{align}
where $Nu$ is the Nusselt number. In our simulation, we consider domain of $129\times 257$ points with $\delta t = 0.3$, $k=0.6$, $C_v=3$, $\nu = 0.005$, $\kappa = 0.0234$ and $g = 0.0001$ in lattice units. The non-dimensional numbers are $Ja=0.064, Pr = 0.094$ and $Gr=2482.58$. Based on the value of the Jacob number, the Nusselt number as computed from the correlation \eqref{eq:klimenko1} amounts to $Nu_k = 2.6085861$. 
Initially, the liquid-vapor interface is perturbed with the function
\begin{align}
    y = 0.125W-0.05W \cos \left(\frac{2\pi x}{W}\right),
\end{align}
where $W$ is the width of the domain. The space-averaged Nusselt number is computed throughout the simulation using
\begin{align}
    \langle Nu \rangle = -\frac{l_s}{W(T_w-T_{sat})} \int_0^W \frac{\partial T}{\partial y}\bigg|_w \, dx,
\end{align}
where the gradient of the temperature is computed at the wall using finite differences. 
The evolution of the liquid-vapor interface is shown at three different times in Fig. \ref{fig:film-boiling-contour}. 
The first bubble is released at $t^*\approx 15$, which is then followed by a periodic release of bubbles. The space-averaged Nusselt number is computed during the simulations until the first bubble is released. The results are presented in Fig. \ref{fig:film-boiling-Nusselt}, where we have compared the time-averaged Nusselt number with the correlation \eqref{eq:klimenko1}. We can confirm the time-averaged Nusselt number is in very close agreement with the correlation while according to \cite{klimenko1982film}, \red{the majority of the experimental data lie within $\pm 25\%$ interval of the fitted lines obtained by Eqs. \eqref{eq:klimenko1}-\eqref{eq:klimenko2}}.
    
\begin{figure}[ht]
  \centerline{\includegraphics[width=1\textwidth]{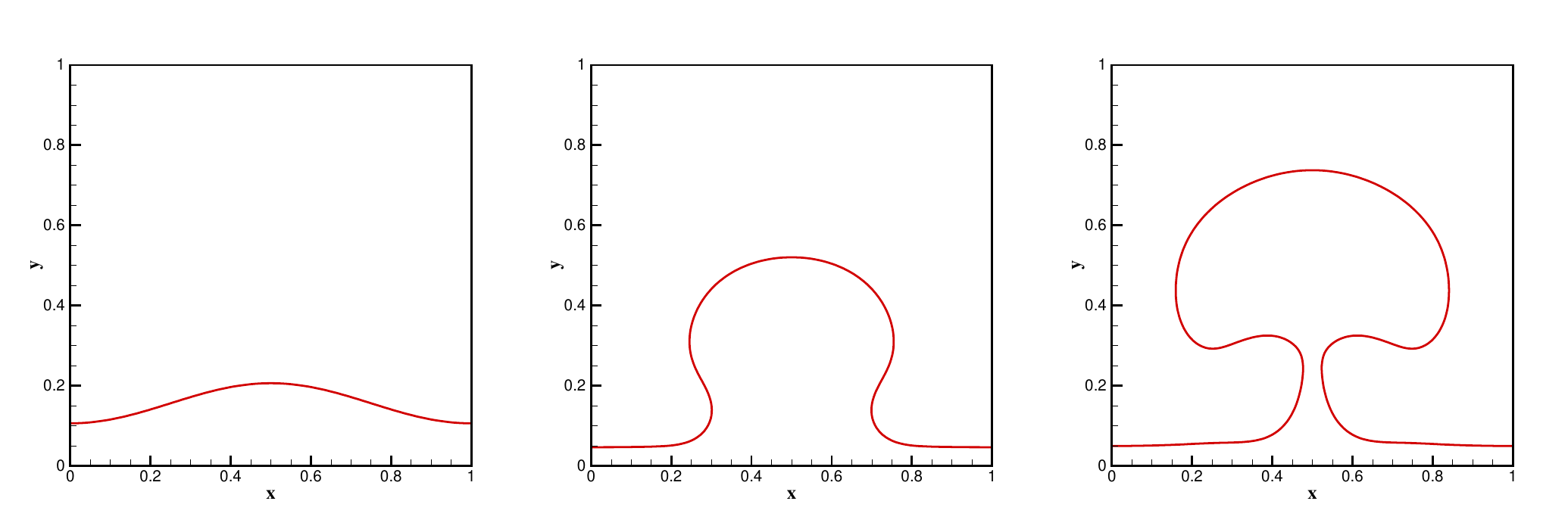}}
  \caption{Bubble growth from the vapor film at $Gr = 2482.58$, $Ja=0.064$ and $Pr=0.094$. The phase boundary is shown at different times. From left to right: $t^*=0$, $t^*=9.8$, $t^*=14.96$}
\label{fig:film-boiling-contour}
\end{figure}
\begin{figure}[ht]
  \centerline{\includegraphics[width=1\textwidth]{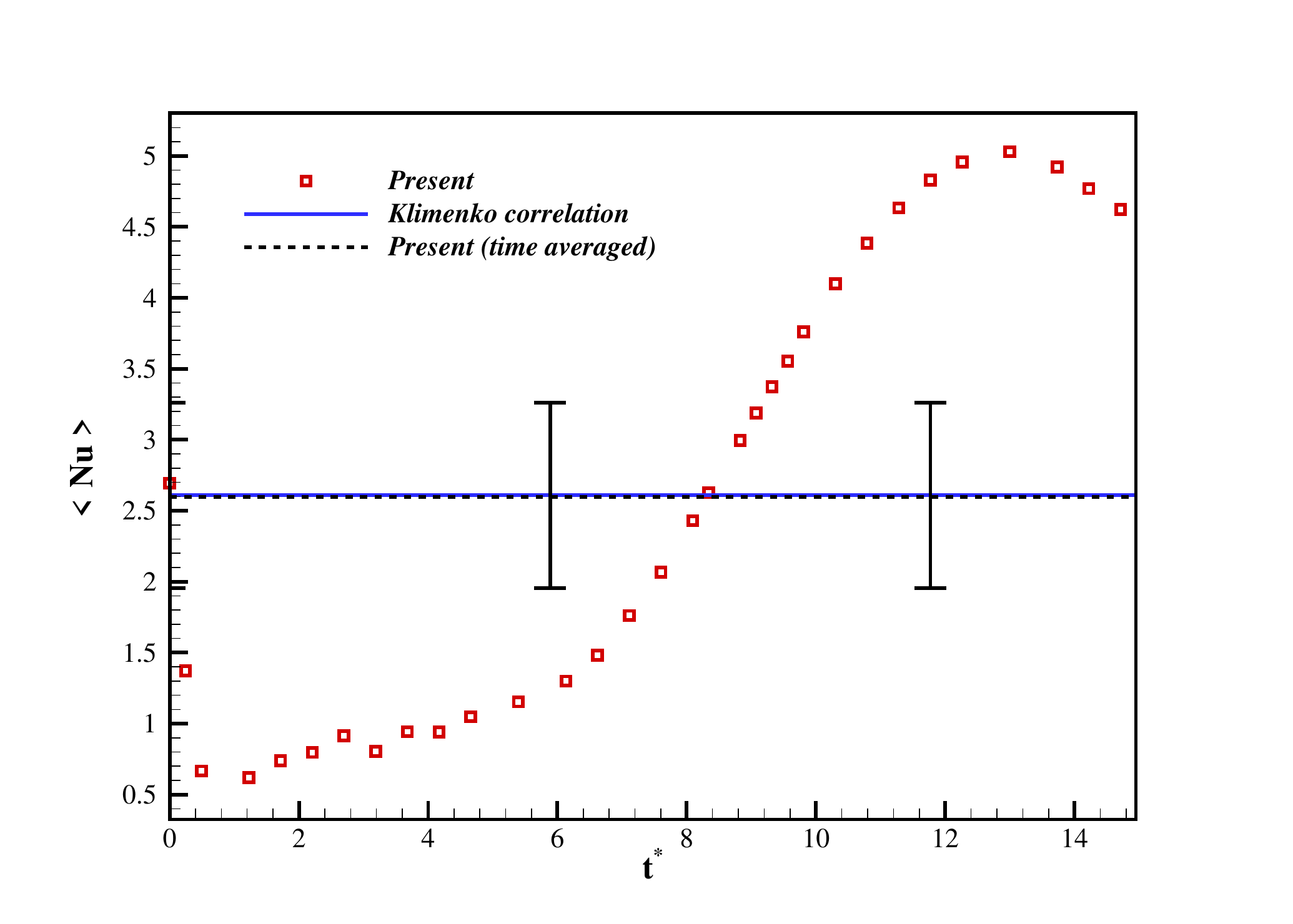}}
  \caption{Space-averaged Nusselt number as a function of dimensionless time for $Gr = 2482.58$, $Ja=0.064$ and $Pr=0.094$. The error bars amount to $\pm 25\%$ acceptable error as shown by Klimenko \cite{klimenko1982film}.}
\label{fig:film-boiling-Nusselt}
\end{figure}
\subsection{On the stability of the shock waves}
The stability of planar shock waves subject to small perturbations have long been investigated since the pioneering work of D'yakov \cite{dyakov1954shock} and later modifications of Kontorovich \cite{kontorovich1958}, which were the first attempts to study the conditions under which a planar shock with corrugations on its surface would become unstable. The key parameter in the analysis of shock instabilities is the so called D'yakov parameter \cite{bates-prl2000}, defined as 
\begin{align}
    h_D = j^2 \left( \frac{dv}{dp} \right)_H,
    \label{eq:dyakov parameter}
\end{align}
evaluated at the post-shock (downstream) state, where $j^2 = (p_1 - p_0)/(v_0 - v_1)$ is the square of the mass flux across the shock front, $v$ is the specific volume, $p$ is the thermodynamic pressure and the subscript $H$ denotes that the derivative is taken along the Hugoniot curve \cite{bates1999some}.

Furthermore, the subscripts "0" and "1" refer to the upstream (pre-shock) and downstream (post-shock) states, respectively. It has been shown that the necessary condition for stability of a shock wave is \cite{bates2007instability,bates2012jfm}:
\begin{align}
    -1<h_D<1+2M_1,
    \label{eq:stability}
\end{align}
where $0<M_1<1$ is the downstream Mach number, which is measured in the reference frame that is moving with the shock. Under this condition, linear perturbations imposed on the shock front will asymptotically decay in time as $t^{-3/2}$ \cite{bates2012jfm}.

According to the theory, a planar shock wave is unconditionally stable
when propagating through an ideal-gas medium \cite{landau1987fluid}. This can be easily evaluated, where $h$ for an ideal gas EOS yields $h_{D,\rm ig}=-1/M_0^2$, which always falls within the stability range \eqref{eq:stability}. On the other hand, for non-ideal fluids, these stability conditions \eqref{eq:stability} can be violated, which  leads to an amplification of the perturbations until the structure of the flow filed is altered \cite{bates2012jfm}. It has been shown that the violation of the upper limit of the stability condition \eqref{eq:stability} corresponds to the splitting of the shock front into two counter-propagating waves \cite{gardner1963comment}, while the violation of the lower limit is associated with the splitting of the shock front into two waves, travelling in the same direction \cite{bethe1942theory}.

Extensive theoretical investigations have been carried out to study the dynamics of the isolated planar shock waves propagating in an inviscid fluid medium. Namely, Bates \cite{bates2004initial,bates2007instability} derived analytical expressions for the amplitude of the ripples on the shock front, for initial sinusoidal perturbations. According to Ref. \cite{bates2004initial}, two families of solutions emerge depending on the sign of the following non-dimensional parameter:
\begin{equation}
    \Lambda = \alpha^4 - 4\beta\Gamma\alpha^2+4\Gamma^2,
\end{equation}
where
\begin{align}
    \alpha^2=\frac{1-M_1^2}{M_1^2}, \ \beta=\frac{1-h_{D_1}}{2M_1}, \ \Gamma=\frac{(1+h_D)\eta}{2M_1}
\end{align}
are non-dimensional parameters as a function of downstream conditions and $\eta=\rho_1/\rho_0$ is the compression ratio through the shock. For the case $\Lambda>0$, the solution is given by
\begin{align}
    \frac{\delta x (\tau)}{\delta x(0)} =  \frac{2\alpha^2\sqrt{\beta^2-1}}{\sqrt{\Lambda}} \int_0^{\infty}&(b\sin{az}\cos{bz}-a\cos{az}\sin{bz}) 
    \frac{J_1(\alpha(\tau+z)}{\alpha(\tau+z)}dz,
    \label{eq:ripple}
\end{align}
where $\delta x$ is the amplitude of the ripple, $\tau=Ukt/\eta$ is the non-dimensional time, $U$ is the speed of the shock front in the laboratory reference of frame, $k$ is the wave number of the initial ripple, $J_1$ is the first-kind Bessel function and the parameters $a$ and $b$ are defined as
\begin{align}
a &= \sqrt{\frac{2\beta\Gamma-\alpha^2}{4(\beta^2-1)}+\frac{\Gamma}{2(\beta^2-1)^{1/2}}} ,\\
b &= \sqrt{\frac{2\beta\Gamma-\alpha^2}{4(\beta^2-1)}-\frac{\Gamma}{2(\beta^2-1)^{1/2}}} ,
\end{align}
and are real numbers. It is interesting to mention that the ideal-gas EOS belongs to this class of solutions. Using the asymptotic approximation $J_1(x) \sim \sqrt{2}(\pi x)^{-1/2}\cos{(x-3\pi/4)}$ as $x\to\infty$ and considering Eq. \eqref{eq:ripple}, one can confirm that the amplitude of the ripple in a fluid medium with $\Lambda>0$ (such as an ideal gas) will decay in time with the negative power law $\tau^{-3/2}$ in the long-time limit \cite{bates2004initial}.

However, the situation can be different for fluids with an \red{EOS} that \red{can yield} a negative $\Lambda$. Finally, for the case $\Lambda<0$, the solution to the initial value problem is \cite{bates2004initial}:
\begin{align}
\frac{\delta x (\tau)}{\delta x(0)} &=  
\frac{1}{2}\exp(-\sigma\tau)\cos{a\tau}-\left[  \frac{\Gamma}{\beta^2-1}-\frac{\alpha^2}{2(\beta^2-1)} \right ]\frac{\exp(-\sigma\tau)\sin{a\tau}}{4a\sigma}\nonumber\\
& + \frac{\alpha^2}{4\sigma\sqrt{\beta^2-1}}
\Bigg\{ \int_0^\tau \exp\left(-\sigma(\tau-z)\right)\left(\cos{a(\tau-z)}+\frac{\sigma}{a}\sin{a(\tau-z)}\right)\frac{J_1(\alpha z)}{\alpha z} dz \nonumber\\
& + \int_0^\infty \left(\cos{az}+\frac{\sigma}{a}\sin{az}\right)\frac{J_1(\alpha(\tau+z))}{\alpha(\tau+ z)} dz \Bigg\},
\label{eq:ripple2}
\end{align}
where $\sigma=-ib$ is a real number. The presence of the exponential function implies a stronger damping compared to Eq. \eqref{eq:ripple}. However, the long time asymptotic is still a function of $\tau^{-3/2}$ in both cases \cite{bates2004initial}.

These theoretical considerations give us the opportunity to test and validate our numerical model also in the high-speed regime for the exotic shock-wave behavior of non-ideal gases.
Our simulations consist of a long channel with periodic boundary conditions in the vertical direction. In all cases, the conductivity is set to zero and the viscosity is chosen to take the lowest possible value as long as the simulations are stable. In order to capture the shocks and avoid oscillations at the shock front, a third-order WENO scheme based on a 4-point stencil has been used in the reconstruction process instead of the third-order Lagrange polynomials in all previous simulations. 
Three different cases have been selected; ideal gas (IG) with $M_0=3$, vdW fluid with $M_0=3.033$ and $M_0=1.114$. All other parameters are provided in Table \ref{tab:table1}. In all cases, the shock front is initially perturbed with a single-mode sinusoidal function. The ratio of the amplitude to the wave length of the perturbation is $10\%$. The first two cases fall into the category of stable shocks, where the perturbations on the shock front are expected to decay in time. However, the last case is an example of shock-splitting, which will be discussed below. 

Let us consider the first two cases. At time $t=0$, the shock starts to propagate while it oscillates as it moves further towards the low-pressure side. 
We then measure the oscillation amplitude and compare it to the analytical expressions.

\begin{table}[H]
\caption{\label{tab:table1}Parameters for different cases of the simulation of the shock-stability}
\centering
\begin{tabular}{cccccccccc}
\toprule
 Case & EOS & $M_0$ & $\rho_0$ & $p_0$& $\nu$ & $\delta t$ & $C_v/R$ & $h_D$ & $\Lambda$ \\
\midrule
    (1) & IG & 3.0 &  1 & 1 & $10^{-3}$ & 0.03 & 1.5 & -1/9 & 4.214\\
    (2) & vdW & 3.033 &  $\rho_{cr}/3$ & $0.66p_{cr}$ & $10^{-4}$ & 0.1 & 3.0 & -0.094  & -7.856\\
    (3) & vdW & 1.114 & $\rho_{cr}/3$ & $0.66p_{cr}$ & $10^{-6}$ & 0.4 & 80.0 & -0.542  & 1.487\\
\bottomrule
\end{tabular}
\end{table}
The initial shape of the shock front and its evolution in time is shown in Fig. \ref{fig:ripple-front} for the first case, where the oscillation of the front and the damping effect is apparent. Figure \ref{fig:perturbation-case1} shows the bird-eye view of this simulation.

According to the sign of $\Lambda$, the magnitude of the ripple for case (1) and case (2) were compared with analytical solutions \eqref{eq:ripple} and \eqref{eq:ripple2}, respectively. The results are presented in Fig. \ref{fig:ripple} and are in good agreement with the theory. It is apparent that our model captured the two distinct damping effects accurately, with more pronounced damping for the non-ideal fluid, as expected.

We now consider the exotic case (3). As mentioned earlier, non-ideal fluids can show exotic behaviors under certain conditions. Regarding case (3), due to its large specific heat value, the Hugoniot curve passes through regions, where the relation $h_D<-1$ is satisfied. 
As argued in \cite{bates1999some}, this, together with the fact that the Hugoniot curve has more than two intersection points with the Rayleigh line (see Fig. \ref{fig:hugoniot}), can cause the shock front to split into two traveling waves. Figure \ref{fig:shock-front-split} presents our simulation for this case, where it is visible that the initial perturbation on the shock front has become unstable, leading to splitting of the shock. 
The resulting waves, travel in the same direction with different speeds, as expected. This validates our model also for high-speed flows and shock-waves in real-gas media.\\ 
\begin{figure}
  \centerline{\includegraphics[clip, trim = 0 2cm 0 1cm ,width=0.6\textwidth]{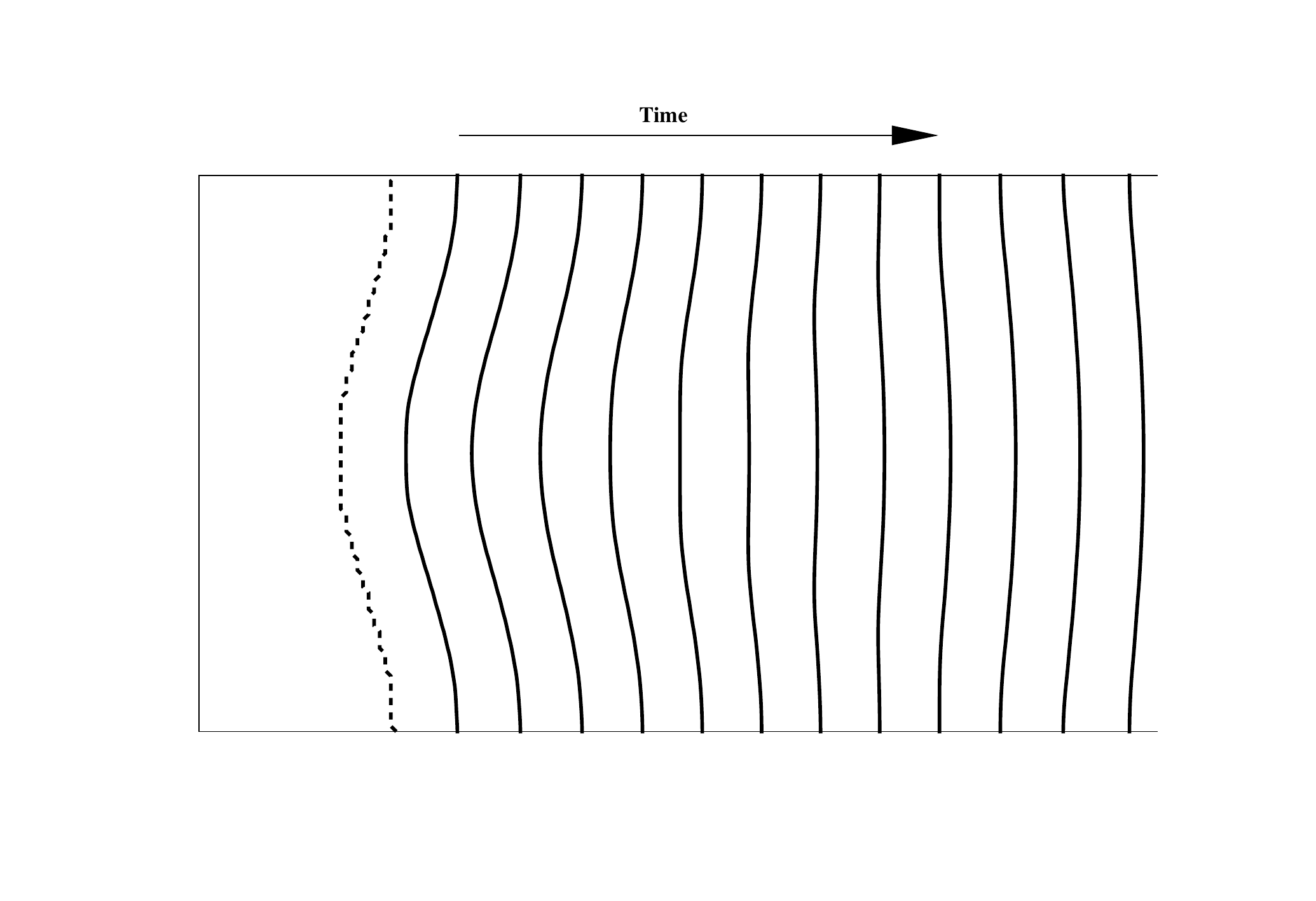}}
  \caption{Evolution of an initially perturbed shock (dashed line) in time in an ideal gas medium with $\gamma=5/3$ and $\rm{Ma}=3.0$.}
\label{fig:ripple-front}
\end{figure}
\begin{figure}[ht]
  \centerline{\includegraphics[width=1\textwidth]{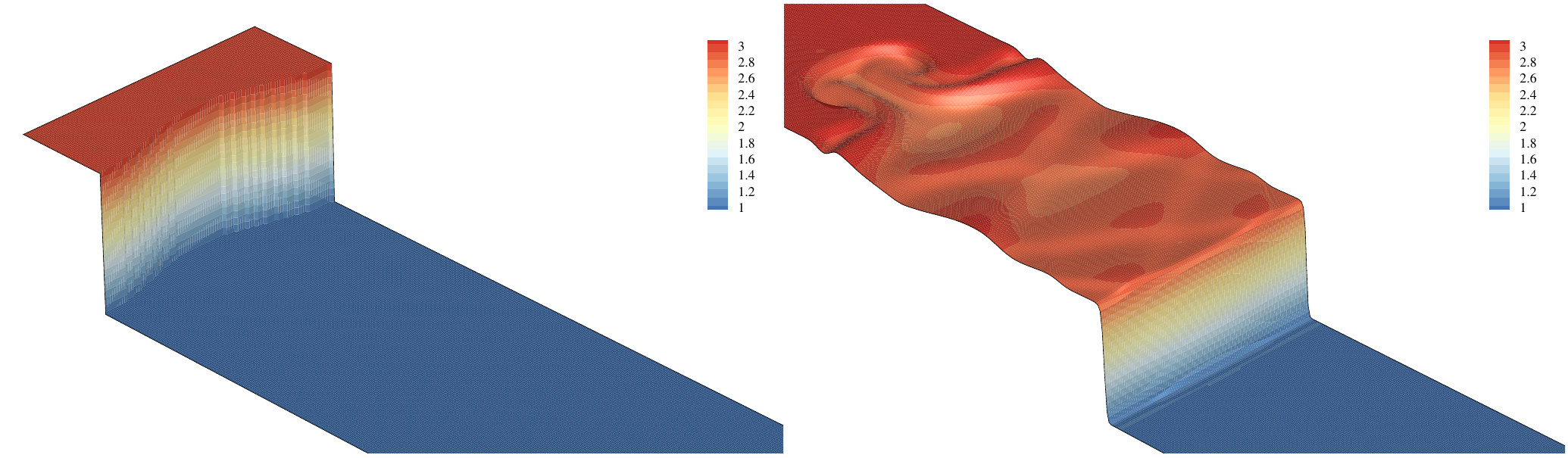}}
  \caption{Case(1): Left: Initial perturbation on the shock front. Right: evolution of the shock-front at time $\tau = 9.73$. Both plots and their coloring, show reduced density with respect to the pre-shock value $\rho/\rho_0$.}
\label{fig:perturbation-case1}
\end{figure}
\begin{figure}[ht]
  \centerline{\includegraphics[width=\textwidth]{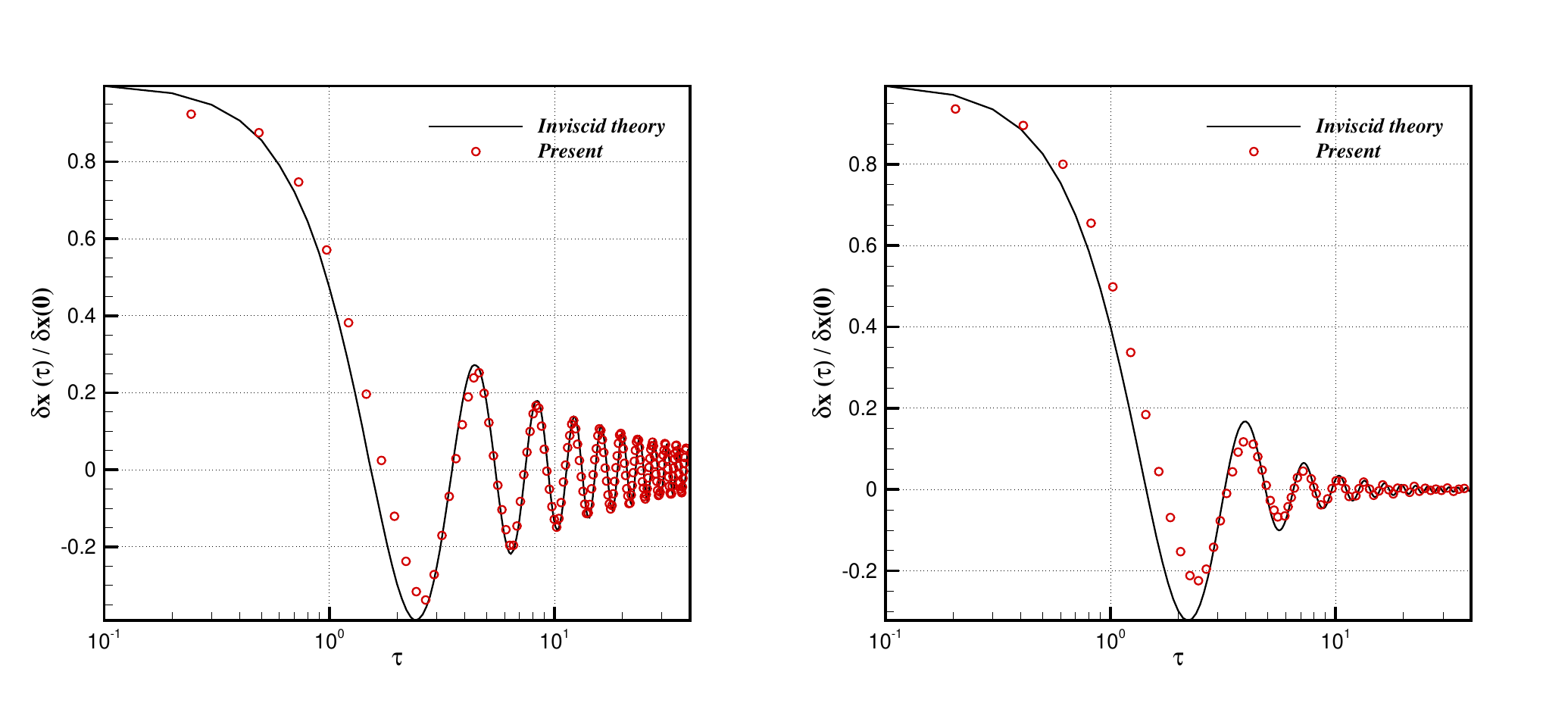}}
  \caption{Comparison between the theoretical solution and the simulations for the ripple amplitude of an initially perturbed shock propagating through (left) an ideal gas with $M_0=3.0$ (right) a vdW fluid with $M_0=3.033$.}
\label{fig:ripple}
\end{figure}
\begin{figure}[ht]
  \centerline{\includegraphics[width=0.8\textwidth]{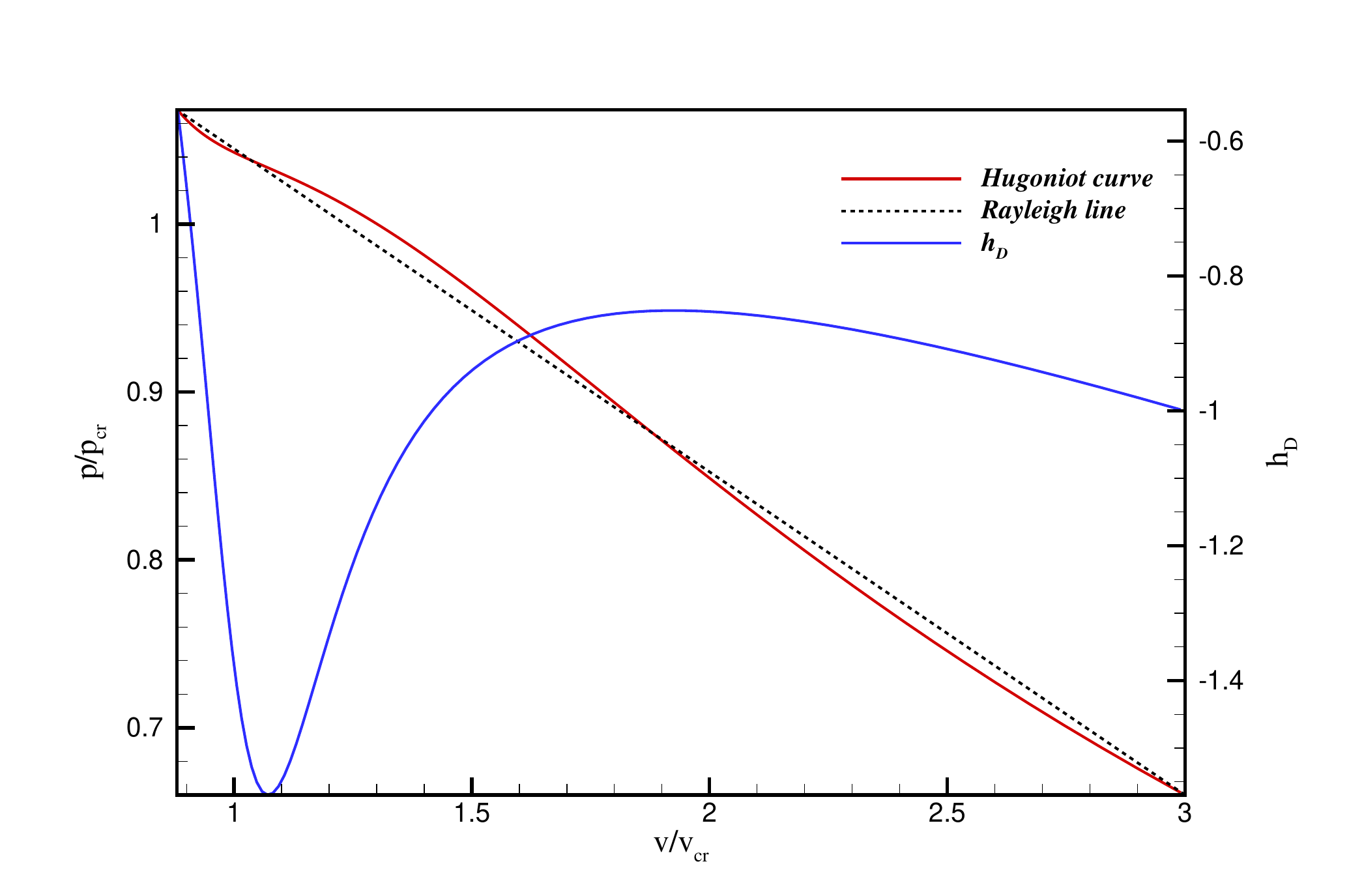}}
  \caption{Case(3): The plot of $h_D$ and the Hugoniot curve as a function of the downstream specific volume. It is visible that the Hugoniot curve has more than two intersection points with the Rayleigh line. Also, as the volume decreases, there are regions where $h_D<-1$.}
\label{fig:hugoniot}
\end{figure}
\begin{figure}[ht]
  \centerline{\includegraphics[width=1\textwidth]{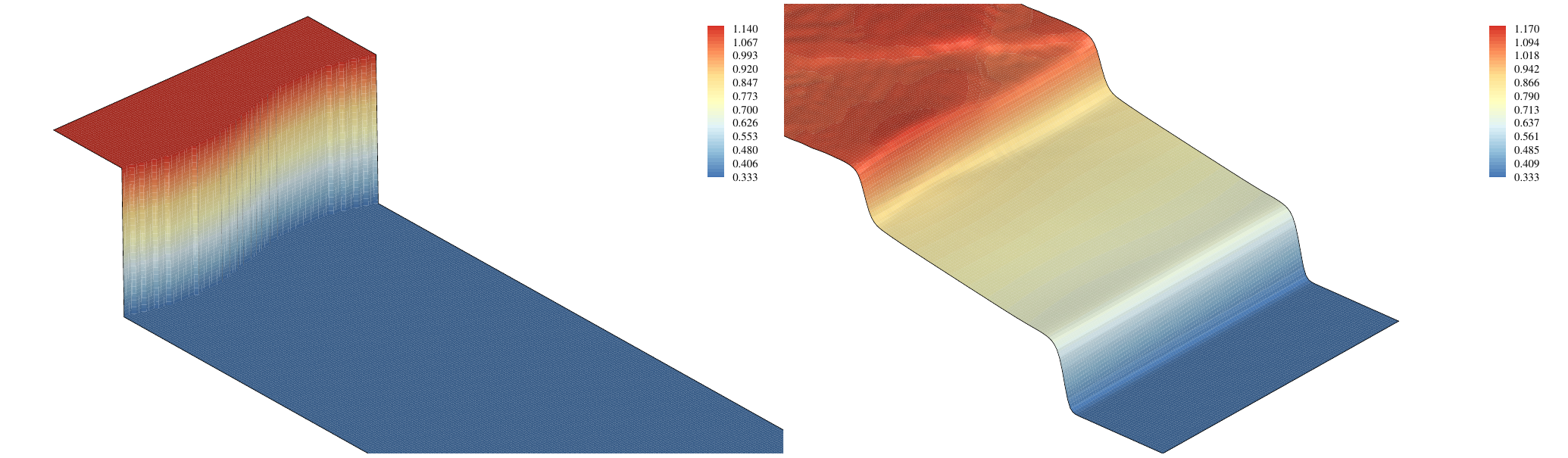}}
  \caption{Case(3): Left: initial perturbation on the shock front. Right: Evolution of the shock-front at time $\tau = 16.3$. The shock wave has split into two travelings waves in the same direction. Both plots and their coloring show reduced density with respect to the critical value $\rho/\rho_{cr}$.}
\label{fig:shock-front-split}
\end{figure}
\section{Conclusion}\label{sec:conclusion}
In this paper, we have presented a thorough study of our recently proposed model for compressible non-ideal flows.
The model features full Galilean-invariance and the full energy equation is recovered for a non-ideal fluid, accounting also for  two-phase systems and the presence of interfaces.
It has been shown that the model is able to handle flows which are far into supercritical states. The effect of the inversion line on the $T-P$ diagram was correctly captured for the van der Waals fluid in a wide range of reduced-pressure\red{s}; from $p_r=3$ to $p_r=15$. In addition, owing to the full energy conservation, the latent heat is already included in the model. This was shown on two different phase-change benchmarks: The one-dimensional Stefan problem and \red{b}oiling. As one of the advantages of the model, we were able to choose relatively large Stefan and Jacob numbers, which are scarce in the literature. 
Finally, the stability of an initially perturbed shock front in ideal gas and in the van der Waals fluid \red{at a} Mach number $\rm{Ma} \approx 3$ were studied and compared to theoretical predictions. It was observed that the damping effect is much stronger in the nonideal fluid as predicted by the inviscid theory. Beside from the fact that all of these simulations were implemented by taking only nine  discrete velocities in two dimensions, the results show that the real-gas effects are captured accurately by the proposed model.

\vspace{6pt}

\authorcontributions{E.R., B.D. and I.K. conceptualized the model. E.R. and B.D. developed the code. E.R. ran the simulations and validated the results. I.K. supervised the project. All authors contributed to writing the paper.}

\funding{This work was supported by the European Research Council (ERC) Advanced Grant No. 834763-PonD and the Swiss National Science Foundation (SNSF) Grant No. 200021-172640.
Computational resources at the Swiss National Super Computing Center (CSCS) were provided under Grant No. s897.}

\acknowledgments{The authors would like to thank Prof. Jason Bates who provided useful information regarding his studies on shock stability analysis.}

\conflictsofinterest{The authors declare no conflict of interest.} 

\abbreviations{The following abbreviations are used in this manuscript:\\

\noindent 
\begin{tabular}{@{}ll}
LBM & Lattice Boltzmann method\\
EOS & Equation of state\\
PonD & Particles on Demand
\end{tabular}}

 \reftitle{References}
 \externalbibliography{yes}
 \bibliography{References}

\end{document}